%% file: main.tex
\def\NOTES{0}
\def\TR{1}

\if \TR 1 
 \documentclass[journal]{IEEETran}
\else 
 \documentclass[10pt]{sig-alternate-05-2015}
\fi

\usepackage[utf8]{inputenc}
\usepackage[T1]{fontenc}

\usepackage{times}
\usepackage[hidelinks]{hyperref} 
\hypersetup{pdfstartview=FitH,pdfpagelayout=SinglePage}

\usepackage{amsmath}
\usepackage{amsmath}
\usepackage{amssymb}
\usepackage{booktabs}
\usepackage{multirow}
\usepackage{graphicx}
\usepackage{longtable}
\usepackage[normalem]{ulem}
\usepackage{enumitem}
\usepackage{listings}
\usepackage{subfig}
\usepackage{xspace}
\usepackage{nicefrac}
\usepackage{color}
\usepackage{cite}
\usepackage[export]{adjustbox}
\usepackage{bm}
\usepackage{algorithm}
\usepackage{algpseudocode}

\usepackage{tikz}
\usetikzlibrary{calc,positioning,arrows,chains,shapes.geometric}

\if \NOTES 1
\newcommand{\ed}[1]{\textsf{\textcolor{red}{[mch: #1]}}}
 
\else
 \newcommand{\ed}[1]{}
 
\fi

\newcounter{counter}

\newtheorem{lemma}[counter]{Lemma}
\newtheorem{theorem}[counter]{Theorem}

\newcommand{\nphard}{NP-hard}
\newcommand{\obliviousperdestination}{\textsc{Oblivious IP routing}\xspace}
\newcommand{\obliviousperdestinationdag}{\textsc{Oblivious IP routing}\xspace}
\newcommand{\bipartition}{\textsc{Bipartition}\xspace}
\newcommand{\integergadget}{\textsc{Integer}\xspace}

\newcommand{\defaultECMPDAG}{default-ECMP-DAG\xspace}
\newcommand{\localSearchECMPDAG}{local-search-ECMP-DAG\xspace}
\newcommand{\augmented}{-augmented\xspace}
\newcommand{\ecmp}{ECMP\xspace}
\newcommand{\eat}[1]{}

 \makeatletter
 \newcommand*{\fnsymb}[1]{%
   \ensuremath{%
     \ifcase#1
     \or 
       *%
     \or 
       \dagger
     \or 
       \ddagger
     \or 
       \mathsection
     \or 
       \mathparagraph
     \else 
       \@ctrerr
     \fi
   }%
 }
 \makeatother

\begin{document}


\if \TR 1 
  \title{Lying Your Way to Better Traffic Engineering (Technical Report)}
\else 
  \title{Lying Your Way to Better Traffic Engineering}
\fi

\if \TR 0 
	\numberofauthors{3}%
	\author{
	    \alignauthor%
	    Marco Chiesa\\%
	        \affaddr{Universit\`e catholique de Louvain}%
	        \email{marco.chiesa@uclouvain.be}
	    \alignauthor%
	    Gábor Rétvári\\%
	        \affaddr{MTA-BME Information\\Systems Research Group}%
	        \email{retvari@tmit.bme.hu}
	    \alignauthor%
	    Michael Schapira\\%
	        \affaddr{Hebrew University of Jerusalem}%
	        \email{schapiram@cs.huji.ac.il}
	}
\else 
	\author{\IEEEauthorblockN{Marco Chiesa\IEEEauthorrefmark{1}, G\'abor R\'etv\'ari\IEEEauthorrefmark{2}, Michael Schapira\IEEEauthorrefmark{3}}
		
		\IEEEauthorblockA{\IEEEauthorrefmark{1}Universit\`e catholique de Louvain}
		\IEEEauthorblockA{\IEEEauthorrefmark{2}MTA-BME Information\\Systems Research Group}
		\IEEEauthorblockA{\IEEEauthorrefmark{3}Hebrew University of Jerusalem}
	}
\fi

\date{}

\if \TR 0 
\CopyrightYear{2016} 
\setcopyright{acmlicensed}
\conferenceinfo{CoNEXT '16,}{December 12 - 15, 2016, Irvine, CA, USA}
\isbn{978-1-4503-4292-6/16/12}\acmPrice{\$15.00}
\doi{http://dx.doi.org/10.1145/2999572.2999585}

\clubpenalty=10000 
\widowpenalty = 10000
\fi

\maketitle



\if \TR 1 
	\input{000-tr-abstract}
	\input{100-tr-intro}
	\input{200-tr-motivating-example}

	\input{300-tr-model}

	\input{400-tr-negative-results}
	\input{500-tr-overview-and-design}
	\input{600-tr-evaluation}

	\input{700-tr-prototype}

	\input{800-tr-related-work}

	\input{900-tr-conclusions}
\else 
	\input{000-abstract}
	\input{100-intro}
	\input{300-overview-and-design}
	\input{400-evaluation}
	\input{500-related-work}

	\input{600-conclusions}
\fi

\bibliographystyle{unsrt}
\bibliography{stote}

\if \TR 1

 \input{950-tr-app-closer-look}
\fi

\end{document}

%% file: 000-tr-abstract.tex
\begin{abstract}
To optimize the flow of traffic in IP networks, operators do traffic engineering (TE), i.e.,  tune routing-protocol parameters in response to traffic demands. TE in IP networks typically involves configuring static link weights and splitting traffic between the resulting shortest-paths via the Equal-Cost-MultiPath (ECMP) mechanism. Unfortunately, ECMP is a notoriously cumbersome and indirect means for optimizing traffic flow, often leading to poor network performance. Also, obtaining accurate knowledge of traffic demands as the input to TE is elusive, and traffic conditions can be highly variable, further complicating TE. We leverage recently proposed schemes for increasing ECMP's expressiveness via carefully disseminated bogus information ("lies") to design COYOTE, a readily deployable TE scheme for robust and efficient network utilization. COYOTE leverages new algorithmic ideas to configure (static) traffic splitting ratios that are optimized with respect to all (even adversarially chosen) traffic scenarios within the operator's "uncertainty bounds". Our experimental analyses show that COYOTE significantly outperforms today's prevalent TE schemes in a manner that is robust to traffic uncertainty and variation. We discuss experiments with a prototype implementation of COYOTE.
\end{abstract}

%% file: 100-tr-intro.tex
\section{Introduction}\label{sect:intro}

\noindent{\bf Today's traffic engineering is suboptimal.} To adapt the routing of traffic to the demands network operators do traffic engineering (TE), i.e., tune routing-protocol parameters so as to influence how traffic flows in their networks~\cite{1039866, 4483669, Curtis:2011:DSF:2018436.2018466}. Today's prevalent scheme for TE within an organizational IP network is based on configuring static link-weights into shortest-path protocols such as OSPF~\cite{rfc2328} and splitting traffic between the resulting shortest-paths via ECMP \cite{rfc2992}. Traditional TE with ECMP significantly constrains both route-computation and traffic splitting between multiple paths in two crucial ways: (1) traffic from a source to a destination in the network can only flow along the shortest paths between them (for the given configuration
of link weights), and (2) traffic splitting between multiple paths (if multiple shortest paths exist) can only be done in very specific manners (see Section~\ref{sec:motivating-example} for an illustration). 

ECMP's lack of expressiveness makes traffic engineering with ECMP a notoriously hard task that often results in poor performance. Indeed, not only does ECMP's inflexibility imply that traffic flow might be arbitrarily far from the global optimum~\cite{fortz-thorup:increasing}, but even choosing ``good'' link weights for TE with ECMP is infeasible in general~\cite{chiesa2014traffic}. Beyond ECMP's deficiencies, today's dominant TE schemes also suffer from other predicaments, e.g., obtaining an accurate view of traffic demand so as to optimize TE is elusive, as most networks lack the appropriate measurement infrastructure. Also, traffic can be highly variable and routing configurations that are good with respect to one traffic scenario can be bad with respect to another. We thus seek a TE scheme that is backwards compatible with legacy routing infrastructure (i.e., OSPF and ECMP), yet \emph{robustly} achieves high performance even under uncertain or variable traffic conditions.

\vspace{0.05in}\noindent{\bf SDN to the rescue?} Software-Defined Networking (SDN) comes with the promise of improved network manageability and flexibility. Yet, transition to SDN is extremely challenging in practice as realizing full-fledged SDN can involve drastic changes to the legacy routing infrastructure. Consequently, recent proposals focus on providing ``SDN-like'' control over \emph{legacy} network devices~\cite{Vissicchio:2014:SLL:2670518.2673868,Vissicchio:2015:CCO:2785956.2787497}. However, while such control greatly enhances the expressiveness of today's IP routing, backwards compatibility with legacy equipment and protocols imposes nontrivial constraints on the design of new SDN applications, including that routing be \emph{destination-based} and, typically, absence of an online traffic measurement infrastructure. We explore how ``legacy-compatible SDN control'' can be harnessed to improve TE in traditional IP networks.

\vspace{0.05in}\noindent{\bf COYOTE: optimized, OSPF/ECMP-compatible TE.} We leverage recently introduced approaches for enriching ECMP's expressiveness without changing router hardware\slash software to design COYOTE (COmpatible Yet Optimized TE). Recent studies show that by injecting ``lies'' into OSPF-ECMP (specifically, information about fake links and nodes), OSPF and ECMP can support much richer traffic flow configurations~\cite{Vissicchio:2014:SLL:2670518.2673868,Vissicchio:2015:CCO:2785956.2787497}. We exploit these developments to explore how OSPF-ECMP routing can be extended to achieve consistently high performance even under great uncertainty about the traffic conditions and high variability of traffic. To accomplish this, COYOTE relies on new algorithmic ideas to configure (static) traffic splitting ratios at routers\slash switches that are optimized with respect to \emph{all} (even \emph{adversarially} chosen) traffic scenarios within operator-specified ``uncertainty bounds''. 

Our experimentation with COYOTE on real network topologies shows that COYOTE consistently and robustly achieves good performance even with very limited (in fact, sometimes even no) knowledge about the traffic demands and, in particular, exhibits significantly better performance than (optimized) traditional TE with ECMP. Our experiments with a prototype implementation of COYOTE also demonstrate its performance benefits. We briefly discuss below the algorithmic challenges facing the design of COYOTE and how these are tackled.

As discussed above, we view COYOTE as an important additional step in the recent exploration of how SDN-like functionality can be accomplished without changing today's networking infrastructure (see~\cite{Vissicchio:2014:SLL:2670518.2673868,Vissicchio:2015:CCO:2785956.2787497}). Indeed, COYOTE can be regarded as the \emph{first} legacy-compatible SDN application for TE.

\vspace{0.05in}\noindent{\bf New algorithmic framework: destination-based oblivious routing.} A rich body of literature in algorithmic theory investigates ``(traffic-demands-)oblivious routing''~\cite{Racke:2008:OHD:1374376.1374415, 4032716, Altin:2012:OOR:2345370.2345375}, i.e., how to compute provably good routing configurations with limited (possibly even no) knowledge of the traffic demands. Past studies~\cite{4032716, kulfi} show that, even though lacking accurate information about the traffic demands,  demands-oblivious routing algorithms yield remarkably close-to-optimal performance on real-world networks. Unfortunately, the above-mentioned algorithms involve forwarding packets based on both source and destination and are so inherently incompatible with des\-ti\-na\-tion-based routing via OSPF-ECMP. In addition, realizing these schemes in practice entails either excessive use of (e.g., MPLS) tunneling\slash tagging in traditional IP networks~\cite{4032716,applegate-thorup-03}, or the ubiquitous deployment of per-flow routing software-defined networking infrastructure~\cite{openflow}.

Our design of COYOTE relies on a novel algorithmic framework for demands-oblivious IP routing. We initiate the study of optimizing oblivious routing under the restriction that forwarding is destination-based. In light of the recent progress on enhancing OSPF-ECMP's expressiveness through ``SDN-like'' control, we view the algorithmic investigation of destination-based oblivious routing as an important and timely research agenda. We take the first steps in this direction. Our first result establishes that, in contrast to unconstrained oblivious routing, computing the optimal destination-based oblivious routing configuration is computationally intractable. We show how, via the decomposition of this problem into sub-problems that are easier to address with today’s mathematical toolkit, and by leveraging prior research, good routing configurations can be generated.

\vspace{0.05in}\noindent{\bf Organization.} We motivate our approach to TE via a simple example in Section~\ref{sec:motivating-example} and formulate a major algorithmic challenge facing legacy-compatible TE optimization in Section~\ref{sec:notation-problem}. Our negative results are presented in Section~\ref{sec:challenges}. We present COYOTE's design and explain how it addresses these challenges in Sections~\ref{sec:design}. A more detailed exposition of COYOTE's algorithmic framework is presented in the Appendix. An experimental evaluation of COYOTE on empirically-derived datasets and COYOTE's prototype implementation are presented in Sections~\ref{sec:eval} and~\ref{sec:prototype}, respectively. We discuss related work in Section~\ref{sect:related-work}. We conclude and leave the reader with promising directions for future research in Section~\ref{sec:conc}.



%% file: 200-tr-motivating-example.tex
\section{A Motivating Example}
    \label{sec:motivating-example}

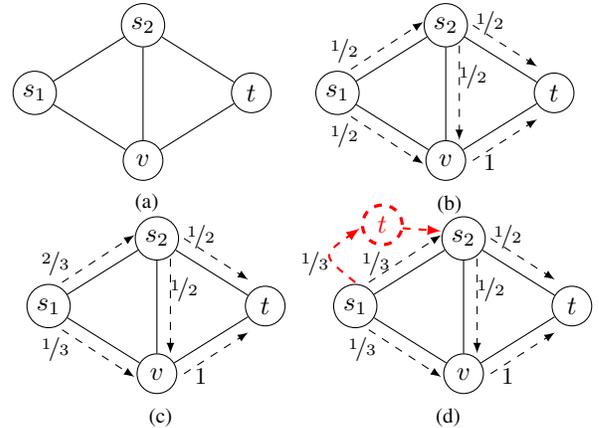
\begin{figure}
  \centering
  \subfloat[][]{%
      \begin{tikzpicture}
        [scale=.18, baseline=(s1.base), minimum size=15,inner sep=2pt, %
        node/.style={anchor=center,circle,draw=black,font=\normalsize}] {
          \node (s1) at (0,0) [node] {$s_1$};%
          \node (s2) at (8,5) [node] {$s_2$};%
          \node (v) at (8,-5) [node] {$v$};%
          \node (t) at (16,0) [node] {$t$};%
        
          \path (s1) edge (s2) edge (v); %
          \path (s2) edge (v) edge (t); %
          \path (v) edge (t); %
        };
      \end{tikzpicture} %
    \label{fig:sample-net}}%
   \hskip1.3em%
    \subfloat[][]{%
      \begin{tikzpicture}
        [scale=.18, baseline=(s1.base), minimum size=15,inner sep=2pt,    %
        node/.style={anchor=center,circle,draw=black,font=\normalsize}] {
          \node (s1) at (0,0) [node] {$s_1$};%
          \node (s2) at (8,5) [node] {$s_2$};%
          \node (v) at (8,-5) [node] {$v$};%
          \node (t) at (16,0) [node] {$t$};%
          
          \path (s1) edge (s2) edge (v); %
          \path (s2) edge (v) edge (t); %
          \path (v) edge (t); %

          \draw[->,>=latex,dashed] ($ (s2) + (1,-1.5) $) -- ($(v)+(1,1.5)$); %
          \node at ($(s2) + (2,-3.8)$) [node,rectangle,draw=none,font=\small] {$\nicefrac{1}{2}$}; %

          \draw[->,>=latex,dashed] ($ (s2) + (2,0) $) -- ($(t) + (-1.25,1.95)$); %
          \node at ($(s2) + (3.25,0.15)$) [node,rectangle,draw=none,font=\small] {$\nicefrac{1}{2}$}; %

          \draw[->,>=latex,dashed] ($ (s1) + (1,1.75) $) -- ($(s2) + (-1.65,.25)$); %
          \node at ($(s1) + (.5,3.15)$) [node,rectangle,draw=none,font=\small] {$\nicefrac{1}{2}$}; %

          \draw[->,>=latex,dashed] ($ (s1) + (1,-1.75) $) -- ($(v) + (-1.65,-0.25)$); %
          \node at ($(s1) + (.5,-3.15)$) [node,rectangle,draw=none,font=\small] {$\nicefrac{1}{2}$}; %

          \draw[->,>=latex,dashed] ($ (v) + (2,0) $) -- ($(t) + (-1.25,-1.95)$); %
          \node at ($(v) + (3.25,-.15)$) [node,rectangle,draw=none,font=\small] {$1$}; %
	};	
      \end{tikzpicture}
    \label{fig:opt-ecmp}}%
  \vspace{-1.5em}%
  \newline%
  \subfloat[][]{%
      \begin{tikzpicture}
        [scale=.18, baseline=(s1.base), minimum size=15,inner sep=2pt,    %
        node/.style={anchor=center,circle,draw=black,font=\normalsize}] {
          \node (s1) at (0,0) [node] {$s_1$};%
          \node (s2) at (8,5) [node] {$s_2$};%
          \node (v) at (8,-5) [node] {$v$};%
          \node (t) at (16,0) [node] {$t$};%
          
          \path (s1) edge (s2) edge (v); %
          \path (s2) edge (v) edge (t); %
          \path (v) edge (t); %

          \draw[->,>=latex,dashed] ($ (s2) + (1,-1.5) $) -- ($(v)+(1,1.5)$); %
          \node at ($(s2) + (2,-3.8)$) [node,rectangle,draw=none,font=\small] {$\nicefrac{1}{2}$}; %

          \draw[->,>=latex,dashed] ($ (s2) + (2,0) $) -- ($(t) + (-1.25,1.95)$); %
          \node at ($(s2) + (3.25,0.15)$) [node,rectangle,draw=none,font=\small] {$\nicefrac{1}{2}$}; %

          \draw[->,>=latex,dashed] ($ (s1) + (1,1.75) $) -- ($(s2) + (-1.65,.25)$); %
          \node at ($(s1) + (.5,3.15)$) [node,rectangle,draw=none,font=\small] {$\nicefrac{2}{3}$}; %

          \draw[->,>=latex,dashed] ($ (s1) + (1,-1.75) $) -- ($(v) + (-1.65,-0.25)$); %
          \node at ($(s1) + (.5,-3.15)$) [node,rectangle,draw=none,font=\small] {$\nicefrac{1}{3}$}; %

          \draw[->,>=latex,dashed] ($ (v) + (2,0) $) -- ($(t) + (-1.25,-1.95)$); %
          \node at ($(v) + (3.25,-.15)$) [node,rectangle,draw=none,font=\small] {$1$}; %
        };
      \end{tikzpicture}
    \label{fig:opt-COYOTE}}%
  \subfloat[][]{%
      \begin{tikzpicture}
        [scale=.18, baseline=(s1.base), minimum size=15,inner sep=2pt,    %
        node/.style={anchor=center,circle,draw=black,font=\normalsize},
        fakenode/.style={anchor=center,draw=red,dashed,very thick,circle,font=\normalsize}, %
        ] {

          \node (fake-t) at (2,6) [fakenode] {\textcolor{red}{$t$}};%

          \node (s1) at (0,0) [node] {$s_1$};%
          \node (s2) at (8,5) [node] {$s_2$};%
          \node (v) at (8,-5) [node] {$v$};%
          \node (t) at (16,0) [node] {$t$};%
          
          \path (s1) edge (s2) edge (v); %
          \path (s2) edge (v) edge (t); %
          \path (v) edge (t); %

          \draw[->,>=latex,dashed] ($ (s2) + (1,-1.5) $) -- ($(v)+(1,1.5)$); %
          \node at ($(s2) + (2,-3.8)$) [node,rectangle,draw=none,font=\small] {$\nicefrac{1}{2}$}; %

          \draw[->,>=latex,dashed] ($ (s2) + (2,0) $) -- ($(t) + (-1.25,1.95)$); %
          \node at ($(s2) + (3.25,0.15)$) [node,rectangle,draw=none,font=\small] {$\nicefrac{1}{2}$}; %

          \draw[->,>=latex,dashed,color=red,thick] ($ (s1) + (0,1.75) $) .. controls ($ (s1) + (-2.5,3.75) $) .. ($(fake-t) + (-1.65,-.25)$); %
          \node at ($(s1) + (-3,3.15)$) [node,rectangle,draw=none,font=\small] {$\nicefrac{1}{3}$}; %

          \draw[->,>=latex,dashed,color=red,thick] ($ (fake-t) + (1.50,-0.25) $) -- ($(s2) + (-1.65,.55)$); %
          \node at ($(s1) + (-3,3.15)$) {}; %

          \draw[->,>=latex,dashed] ($ (s1) + (1,1.75) $) -- ($(s2) + (-1.65,.25)$); %
          \node at ($(s1) + (1.5,3.15)$) [node,rectangle,draw=none,font=\small] {$\nicefrac{1}{3}$}; %

          \draw[->,>=latex,dashed] ($ (s1) + (1,-1.75) $) -- ($(v) + (-1.65,-0.25)$); %
          \node at ($(s1) + (.5,-3.15)$) [node,rectangle,draw=none,font=\small] {$\nicefrac{1}{3}$}; %

          \draw[->,>=latex,dashed] ($ (v) + (2,0) $) -- ($(t) + (-1.25,-1.95)$); %
          \node at ($(v) + (3.25,-.15)$) [node,rectangle,draw=none,font=\small] {$1$}; %
        };
      \end{tikzpicture}
    \label{fig:COYOTE-fibbing}}%
   \caption{A sample network: (a) topology with unit capacity links; (b) 
   per-destination ECMP routing (oblivious performance ratio $\nicefrac{3}{2}$); (c) COYOTE (oblivious performance ratio $\nicefrac{4}{3}$); and (d) COYOTE implementation with a fake node inserted at $s_1$ for realizing the required splitting ratio.}
  \label{fig:COYOTE-overview}
\end{figure}

We next motivate our approach to TE in IP networks through a simple example, which will be used as a running example throughout the sequel.

Consider the toy example in Fig.~\ref{fig:sample-net}. Two network users, $s_1$ and $s_2$, wish to send traffic to target $t$. Suppose that each user is expected to send between 0 and 2 units of flow and each link is of capacity 1. Suppose also that the network operator is oblivious to the actual traffic demands or, alternatively, that traffic is variable and user demands might drastically change over time. The operator aims to provide robustly good network performance, and thus has an ambitious goal: configuring OSPF-ECMP routing parameters so as to minimize link over-subscription across \emph{all} possible combinations of traffic demands within the above-specified uncertainty bounds.

Consider first the traditional practice of splitting traffic equally amongst the next-hops on shortest-paths to the destination (i.e., traditional TE with ECMP, see Fig.~\ref{fig:opt-ecmp}),  where the shortest path DAG towards $t$ is depicted by dashed arrows labelled with the resulting flow splitting ratios. The actual OSPF weights are not needed, in terms of exposition, and are omitted from Fig.~\ref{fig:opt-ecmp}. Observe that if the actual traffic demands are $2$ and $0$ for $s_1$ and $s_2$, respectively, routing as in Fig.~\ref{fig:opt-ecmp} would result in link (over-)utilization that is $\nicefrac{3}{2}$ higher than that of the optimal routing of these specific demands (which can send all traffic without exceeding any link capacity). Specifically, routing as in Fig.~\ref{fig:opt-ecmp} would result in $\nicefrac{3}{2}$ units of traffic traversing link $(v,t)$, whereas the total flow could be optimally routed without at all exceeding the link capacities by equally splitting it between paths $(s_1\ s_2\ t)$ and $(s_1\ v\ t)$. One can actually show that this is, in fact, the \emph{best} guarantee achievable for this network via traditional TE with ECMP, i.e., for \emph{any} choice of link weights, equal splitting of traffic between shortest paths would result in link utilization that is $\nicefrac{3}{2}$ higher than optimal for \emph{some} possible traffic scenario. Can we do better?

We show that this is indeed possible if more flexible traffic splitting than that of traditional TE with ECMP is possible. One can prove that for \emph{any} traffic demands of the users, per-destination routing as in Fig.~\ref{fig:opt-COYOTE} results in a maximum link utilization at most $\nicefrac{4}{3}$ times that of the optimal routing\footnote{In fact, even the routing configuration in Fig.~\ref{fig:opt-COYOTE} is not optimal in this respect. Indeed, COYOTE's optimization techniques, discussed in Section~\ref{sec:design}, yield configurations with better guarantees (see Appendix~\ref{sec:revising-example}).}.

We explain later how COYOTE realizes such uneven per-destination load balancing without any modification to legacy OSPF-ECMP. We next formulate the algorithmic challenge facing COYOTE's design.

%% file: 300-tr-model.tex
\section{The Algorithmic Challenge}\label{sec:notation-problem}

Recent proposals advocate ``SDN-like'' control over \emph{legacy} network devices~\cite{Vissicchio:2014:SLL:2670518.2673868,Vissicchio:2015:CCO:2785956.2787497}. By carefully crafting ``lies'' (fake links and nodes) to inject into OSPF-ECMP , OSPF and ECMP can be made to support much richer traffic flow configurations. We aim to investigate how these recent advances can be harnessed to improve TE in traditional IP networks. 

Importantly, while the proposed SDN approach to legacy networks discussed above can greatly enhance the expressiveness of today's IP routing, compatibility with legacy equipment and routing protocols induces nontrivial constraints on the design of ``legacy-compatible SDN applications'': (1) that routing be \emph{destination-based}, and (2) typically, the absence of an online traffic measurement infrastructure. Algorithmic research on traffic flow optimization, in contrast, almost universally allows the routing of traffic to be dependent on both sources and targets, and often involves accurate and up-to-date knowledge of the traffic demand matrices.

We thus seek an algorithmic solution to the following natural and, to the best of our knowledge, previously unexplored, challenge: Compute \emph{destination-based} (i.e., IP-routing-compatible) routing configurations that optimize the flow of traffic with respect to operator-specified ``uncretainty margins'' regarding the traffic demand matrices. We next proceed to formulating this challenge. Our model draws upon the ideas presented in~\cite{4032716}.

\vspace{.1in}
\noindent\textbf{Network, routing, and traffic splitting. } The network is modeled as a directed and capacitated graph $G=(V,E)$, where $c_e$ denotes the capacity of edge $e$. A \emph{routing configuration} $\phi$ on network $G$ specifies, for each vertex $t \in V $, and for each edge $e=(u,v) \in E$, a value $\phi_{t}(e)$ representing the fraction of the flow to $t$ entering vertex $u$ that should be forwarded through edge (link) $e$. Clearly, for every vertex $v$, $\sum_{(u,v)} \phi_t(u,v) = 1$. Observe that the combination of all $\phi_{t}(e)$ values (across all vertices $t$ and edges $e$) indeed completely determines how flow will be routed between every two end-points. 

Since routing is required to be destination-based, the routes to each destination vertex must form a directed acyclic graph (DAG). This is formally captured by requiring that for every vertex $t\in V$ and directed cycle $C$ in $G$, for some edge $e\in C$ on the cycle $\phi_t(e)=0$. We say that a routing configuration $\phi$ that satisfies this condition is a {\em per-destination} (PD) routing configuration. For a PD routing configuration $\phi$, let $f_{st}(v)$, for vertices $s,t,v \in V$, be the fraction of the demand $s \rightarrow t$ that enters $v$. Observe that in PD routing, $f_{st}(v)$ is well-defined and is induced by the $\phi_{t}(e)$ values as follows: $f_{st}(v)=\sum_{e=(u,v)} f_{st}(u) \phi_t(e)$ if $v \neq s$, $1$ otherwise. Observe that when $x$ units of flow are routed from $s$ to $t$ through the network, the contribution of this flow to the load on link $(u,v)$ is $xf_{st}(u)\phi_t(u,v)$. 


\vspace{.1in}
\noindent\textbf{Performance ratio.} We are now ready to formalize the optimization objective. Our focus is on the traditional goal of minimizing link (over-)utilization (often also called ``congestion'' in TE literature). Given a Demand Matrix (DM) $D=\{d_{s_1t_1},\dots,d_{s_kt_k}\}$ specifying the demand between each pair of vertices, the {\em maximum link utilization} induced by a PD routing $(\phi,f)$ is
\begin{displaymath}
MxLU(\phi,D)=\max_{e=(u,v)\in E}\nicefrac{\sum_{s,t\in D}d_{st}f_{st}(u)\phi_t(e)}{c_{e}}\text{.} 
\end{displaymath}

An {\em optimal} routing for $D$ is a PD routing that minimizes the load on the most utilized link, i.e.,
\begin{displaymath}
OPTU(D)=\min_{\phi |\phi \text{ is a PD routing}} MxLU(\phi,D)\text{.}
\end{displaymath}

The {\em performance ratio} of a given PD routing $\phi$ on a specific given DM $D$ is $PERF(\phi,\{D\})=\nicefrac{MxLU(\phi,D)}{OPTU(D)}$. Given a set ${\cal D}$ of DMs, the performance ratio of a PD routing $\phi$ on ${\cal D}$ is $PERF(\phi,{\cal D})=\max_{D\in {\cal D}}PERF(\phi,\{D\})$. ${\cal D}$, in this formulation, should be thought of as the space of  demand matrices deemed to be feasible by the network operator. When ${\cal D}$ is the set of \emph{all} possible DMs, the performance ratio is referred to as the {\em oblivious} performance ratio.
%
%
%
A routing $\phi$ is {\em optimal} if $PERF(\phi,{\cal D})\le PERF(\phi',{\cal D})$ for any $\phi'$. The \obliviousperdestination problem is computing a PD routing $\phi$ that is optimal with respect to a given set ${\cal D}$ of DMs. 
The \obliviousperdestination can be formulated as the following non-linear minimization problem. 
\begin{align}
& \min  \  \alpha \nonumber\\
&  \phi \text{ is a PD routing} \nonumber\\
& \forall \text{ edges } e=(u,v): \nonumber \\
& ~~\forall \text{ DMs }D \in {\cal D} \text{ with } OPTU(D)=r: \nonumber  \\
& ~~~~  \nicefrac{\sum_{(s,t)} d_{st}f_{st}(u)\phi_t(e)}{c(e)} \le \alpha r  \nonumber
\end{align}

Observe that this optimization objective thus captures both the computation of per-destination DAGs and the computation of in-DAG traffic splitting ratios. Our focus is on sets of demand matrices ${\cal D}$ that can be defined through linear constraints as such sets are expressive enough to model traffic uncertainty, but mitigate the complexity of the optimization problem. Specifically, the actual demand $d_{st}$ from a vertex $s$ to $t$ can assume any value in the range $d_{st}^{min} \le d_{st} \le d_{st}^{max}$, where $d_{st}^{min}$ and $d_{st}^{max}$ are the operator's ``uncertainty margins'' regarding $d_{st}$ and are given as input.

%% file: 400-tr-negative-results.tex
\section{Negative Results}
\label{sec:challenges}

We formulated, in Section~\ref{sec:notation-problem}, the fundamental algorithmic challenge facing COYOTE's design: optimizing (traffic-demands-)oblivious routing in IP networks. Importantly, this optimization goal is closely related to the rich body of literature in algorithmic theory on ``unconstrained'' (i.e., not limited to destination-based) oblivious routing~\cite{Racke:2008:OHD:1374376.1374415, 4032716, Altin:2012:OOR:2345370.2345375}.

Our results in this section show that imposing the real-world limitation that routing be destination-based renders the computation of ``good'' oblivious routing solutions significantly harder. We first show that, in contrast to unconstrained oblivious routing~\cite{4032716}, destination-based oblivious routing is intractable, in the sense that computing the optimal routing configuration is NP-hard. Worse yet, in general, the oblivious performance ratio, i.e., the distance from the best demands-aware routing solution can be very high (as opposed to logarithmic for unconstrained oblivious routing). We explain in Section~\ref{sec:design} how COYOTE's design addresses these obstacles. We next present our two negative results.

\subsection{Oblivious IP Routing is NP-Hard}\label{sec:np-hard}

We examine the computational complexity of the \obliviousperdestination problem, as formulated in Section~\ref{sec:notation-problem}. We present the following computational hardness result.\vspace{.1in}

\newcommand{\theohardness}{The \obliviousperdestination problem is \nphard\ even if ${\cal D}$ consists of only two possible demand matrices, only two vertices can source traffic, and all traffic is destined for a single target vertex. }

\begin{theorem}\label{thm:np-complete}
\theohardness
\end{theorem}

\vspace{.1in}
\noindent\textbf{Proof of Theorem~\ref{thm:np-complete}. }
Our proof reduces the \bipartition problem to \obliviousperdestination. In the \bipartition problem, the input is a set $W=\{w_1,\dots,w_{k}\}$ of $k$ positive integers and the goal is to partition then into two sets $A$ and $B$ such that  the sum of the elements in one partition is equal to the sum of the elements in the other partition.

We now show how to construct an instance $I'$ of the \obliviousperdestination problem from an instance $I$ of the \bipartition problem so that the reduction holds, i.e., $I'$ is a positive instance if and only if $I$ is a positive instance..

\begin{figure}
	\centering
	\begin{minipage}{.3\linewidth}
		\centering
		\includegraphics[width=1\columnwidth]{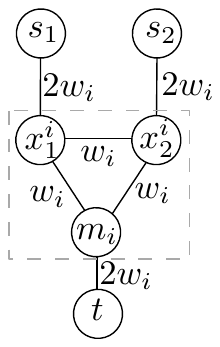}
		\caption{The \integergadget gadget.}
		\label{fig:integer-gadget}
	\end{minipage}
	\hskip1em%
	\begin{minipage}{.6\linewidth}
		\centering
		\includegraphics[width=1\columnwidth]{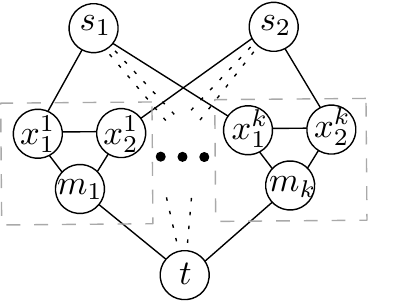}
		\caption{Reduction instance.}
		\label{fig:reduction}
	\end{minipage}
\end{figure}

Let $SUM$ be the sum of all the elements in $W$. We create a directed graph $G$ as follows. We add two source vertices $s_1$ and $s_2$ and a single destination vertex $d$ into $G$. For each integer $w_i$ in $W$, we construct an \integergadget gadget as follows (see Fig.~\ref{fig:integer-gadget}). We add three vertices $x_i^1$, $x_i^2$, and $m_i$. We then add bidirectional edges $(x_i^1,x_i^2)$, $(x_i^1,m_i)$, and $(x_i^2,m_i)$ each with capacity $w_i$. Finally, we add two edges $(s_1,x_i^1)$ and $(s_2,x_i^2)$ with capacity $2w_i$ and edge $(m_i,d)$ with capacity $2w_i$ into $G$.

Observe that we can narrow our attention to demand matrices that can be routed in $G$ without exceeding the edge capacities since the performance ratio is invariant to the rescaling of traffic demands. In addition, as this set describes a convex polyhedron in the demand space, we can further restrict our focus to those vertices of the demand polyhedron that are not ``dominated'' by any other demand vertex, i.e., demand $d=(d_1,\dots,d_n)$ \emph{dominates} demand $d'=(d_1',\dots,d_n')$, if, for all $i=1,\dots,n$, we have $d_i\ge d_i'$.

In our reduction, the min-cut between the source vertices and the target $t$ is $2SUM$, i.e.,  $mincut(s_1,t)=mincut(s_2,t)=mincut(\{s_1,s_2\},t)=2SUM$. 
As such, the set of demand matrices that can be routed within the edge capacities is ${\cal D}=\{(d_{s_1t},d_{s_2t}) | d_{s_1t} + d_{s_2t} \le 2SUM, d_{s_1t}\ge 0, d_{s_2t}\ge 0\}$. As observed in Section~\ref{sec:notation-problem}, the only relevant demand points are $D_1=(d_{s_1t},d_{s_2t})=(2SUM,0)$ and $D_2=(d_{s_1t},d_{s_2t})=(0,2SUM)$, which are vertices of the demands polyhedron. 

There are two crucial routing decisions that have to be made at this point in order to route any demand matrix in ${\cal D}$. The first one is deciding what is the directed acyclic graph that must be used to routed any traffic from the source vertices to $t$. The second one is computing the splitting ratios within that DAG.

Observe that an optimal routing solution for $D_1$ ($D_2$) would orient all the edges $x_i^1$ ($x_i^2$) towards $x_i^2$ ($x_i^1$) in the per-destination DAG rooted at $t$ and split the traffic at $s_1$ in such a way that $2w_i$ units of flow are sent to the $i$'th \textsc{Integer} gadget and equal split is performed at $x_i^1$ ($x_i^2$). In this case, $D_1$ ($D_2$) could be routed without exceeding the edge capacities. This optimal routing for DM $D_1$ ($D_2$) would cause a link utilization of $2$ when routing DM $D_2$ ($D_1$). As such, in order to minimize the oblivious ratio, the crucial routing decision boils down to carefully choose how to orient the edges $(x_i^1,x_i^2)$ in each \textsc{Integer} gadget.

\begin{lemma}\label{lemm:forward-reduction}
	Let $I$ be a positive instance of \bipartition. Then $I'$ has a solution with oblivious performance $\frac{4}{3}$.
\end{lemma}

\begin{IEEEproof}
	Let $(P_1,W \setminus P_1)$ be two equal size partitions of $W$.
	We show how to construct an oblivious routing that has oblivious performance $\frac{4}{3}$.
	
	We define an oblivious routing via splitting ratios at each vertex of the graph, where a splitting ratio of $0$ implies that the outgoing link is oriented in the opposite direction in the per-destination DAG towards $t$.  The splitting ratios $\phi(s_1,x_i^1)$ at $s_1$ ($s_2$) is $\frac{4w_i}{3SUM}$ $\text{ if }w_i \text{ is in } P_1$ ($P_2$), $\frac{2w_i}{3SUM}$ otherwise. The splitting ratios $\phi(x_i^1,x_i^2)$ at $x_i^1$ ($x_i^2$) is $\frac{1}{2}$ $\text{ if }w_i \text{ is in } P_1$ ($P_2$), $0$ otherwise. The splitting ratios $\phi(x_i^1,m_i)$ at $x_i^1$ is $1-\phi(x_i^1,x_i^2) $ and the splitting ratios $\phi(x_i^2,m_i)$ at $x_i^2$ is $1-\phi(x_i^2,x_i^1) $.

	We now show that we can route $D_1$ with congestion at most $\frac{4}{3}$ using the above routing solution.
	Let $C=\frac{4}{3}$.
	Consider an arbitrary integer $w_i\in W$. Two cases are possible: either (i) $w_i$ is in $P_1$ or (ii) not .
	
	In case (i), i.e., $w_i$ is in $P_1$, we send $2SUM\frac{4w_i}{3SUM}=2w_iC$ units of flow to $x_i^1$, which let $(s_1,x_i^1)$ be over-utilized by a factor of $C$. In turn, $x_i^1$ sends  $2w_iC\frac{1}{2}=w_iC$ unit of flow to $m_i$, which let $(x_i^1,m_i)$ be over-utilized by a factor of $C$ and it sends $w_iC$ units of flow to $x_i^1$, which let $(x_i^1,x_i^2)$ be over-utilized by a factor of $C$. The latter flow is forwarded through $(x_i^2,m_i)$ with a link utilization of $C$. Finally, vertex $m_i$ receives two flows, each of $w_iC$ units from $x_i^1$ and $x_i^2$. Since the capacity of edge $(m_i,t)$ is $2w_i$, the link utilization is again $C$.
	
	In case (ii), i.e., $w_i$ is not in $P_1$, we have that a flow of $2SUM\frac{2w_i}{3SUM}=w_iC$ units is sent through edges $(s_1,x_i^1)$, $(x_i^1,m_i)$, and $(m_i,t)$, which causes a link utilization no larger than $C$. 
	
	A similar analysis can be performed to show that the link utilization for DM $D_2$ is never larger than $C$, which proves the statement of the lemma.
\end{IEEEproof}

\begin{lemma}\label{lemm:reversed-reduction}
	Let $I$ be a negative instance of \bipartition. Then $I'$ does not admit a solution with oblivious ratio $\le \frac{4}{3}$
\end{lemma}

\begin{IEEEproof}
	We prove that if $I'$ has an oblivious ratio $\le \frac{4}{3}$, then $I$ is a positive instance for \bipartition. Let $\phi$ be a routing that has oblivious performance ratio $\le \frac{4}{3}$.
	Let $P_1$ be a set of indices such that $i \in P_1$ if $\phi(x_i^2,x_i^1)=0$. Let $P_2=W\setminus P_1$. Two cases are possible: (i) $\sum_{i\in P_1}w_i \le \frac{SUM}{2}$ or (ii) not. 
	
	In case (i), we consider DM  $D_1$, i.e., $(d_{s_1t},d_{s_2t})=(2SUM,0)$. Observe that the maximum amount of flow $F_1$ that can be sent through edges $(x_i^1,m_i)$, with $i=1,\dots,k$ is at most $F_1\le \frac{4}{3}\sum_{i\in P_1}w_i\le \frac{4}{3}SUM$ since the link utilization over all the edges is less than $\frac{4}{3}$. As such, the amount of flow that must be routed through the edges in $\{(x_i^1,x_i^2) | i \in P_1\}$ is at  least $2SUM - F_1\ge 2SUM - \frac{4}{3}SUM = \frac{2}{3}SUM$. This amount of flow is routed without exceeding the edge capacities by a factor higher than $\frac{4}{3}$. This implies that  $\frac{2}{3}\frac{SUM}{\sum_{i\in P_1}{w_i}}\le \frac{4}{3}$, which implies that $\frac{SUM}{2}\le \sum_{i\in P_1}{w_i}$. Since the sum of the element in $P_1$ is no greater than $\frac{SUM}{2}$, the above inequality can be true only if $\sum_{i\in P_1}{w_i}= \frac{SUM}{2}$, that is, $P_1$ is an even bipartition of $I$. Hence, $I$ is a positive instance of \bipartition and the statement of the lemma holds in this case.
	
	In case (ii), i.e., $\sum_{i\in P_1}w_i > \frac{SUM}{2}$, by symmetry, we can apply the same argument used in case (i) to prove that $I$ is a positive instance of \bipartition, where $P_2$ plays the role of $P_1$ and we analyze DM $D_2$ instead of $D_1$. This concludes the proof of the theorem.
\end{IEEEproof}

By Lemma~\ref{lemm:forward-reduction} and~\ref{lemm:reversed-reduction}, the reduction from \bipartition to \obliviousperdestination follows and the statement of Theorem~\ref{thm:np-complete} easily holds.

\subsection{Far-from-Optimal Performance}

Our next negative result shows that in some scenarios even the optimal destination-based oblivious routing can be far from the optimal demands-aware routing (specifically, an $\Omega(|V|)$ factor away, where $V$ is the number of vertices).

\newcommand{\theogap}{There exists a capacitated network graph $G=(V,E)$ and a set ${\cal D}$ of possible traffic matrices such that the performance ratio of the optimal oblivious per-destination routing is $\Omega(|V|)$.}\vspace{0.1in}

\begin{theorem}\label{thm:gap}
\theogap
\end{theorem}
\vspace{0.1in}

\noindent\textbf{Proof of Theorem~\ref{thm:gap}. }
\begin{figure}
\centering
\includegraphics[width=100pt,valign=c]{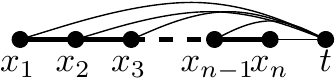}%
\caption{Network example used in the proof of Theorem~\ref{thm:gap}.}
\label{fig:gap-gadget}
\end{figure}
Consider an $n$-vertex path $(x_1,\dots,x_n)$ connected by bidirectional edges with infinite (i.e., arbitrarily high) capacity as in Fig.~\ref{fig:gap-gadget}.
Now, add a destination vertex $t$ and connect each source vertex $x_i$, with $i=1,\dots,n$ to $t$ with a directed edge of capacity $1$. Suppose that the set of possible traffic matrices consists of all possible matrices, i.e., any set of inter-vertex demand values is admissible. We show that the performance of \emph{any} oblivious per-destination routing configuration is $\Omega(\sqrt{n})$.

Consider the set ${\cal D}$ of $n$ possible traffic matrices $D_1, \dots, D_n$ such that in $D_i$ source node $s_i$ wants to send a flow of $n$ unit to $t$ and the demands of all other vertices are $0$ with respect to all target vertices. For each traffic matrix $D_i$, an optimal demands-aware per-destination routing $\phi'$ can route the $x_i\rightarrow t$ flow by carefully splitting it in such a way that each vertex $x_i$ receives a fraction $\nicefrac{1}{n}$ of the flow. Each vertex $x_i$ can then route the received flow directly to $t$ without exceeding the edge capacities, and so $OPTU(D_i)\le1$.
Now, consider any oblivious per-destination routing $\phi$. At least one vertex $x_i$ must send its traffic only through edge $(x_i,t)$, otherwise a forwarding loop exists---a contradiction. This implies that routing $D_i$ with $\phi$ will cause a link utilization of $n$ over edge $(x_i,t)$, i.e., $PERF(\phi,{\cal D})\ge PERF(\phi,D_i)=n$, which concludes the proof of the theorem.

%% file: 500-tr-overview-and-design.tex
\section{COYOTE Design }
\label{sec:design}

\subsection{Overview}

As proved in Section~\ref{sec:challenges}, efficiently computing the \emph{optimal} selection of DAGs \emph{and} in-DAG traffic splitting ratios is beyond reach. We next describe how COYOTE's design addresses this challenge. COYOTE's flow-computation decomposes the task of computing destination-based oblivious routing configurations into two algorithmic sub-problems, and tackles each independently. First, COYOTE applies a heuristic to compute destination-oriented DAGs. Then, COYOTE optimizes in-DAG traffic splitting ratios through a combination of optimization techniques, including iterative geometric programming. We show in Section~\ref{sec:eval} that COYOTE's DAG selection and flow optimization algorithms empirically exhibit good network performance. 

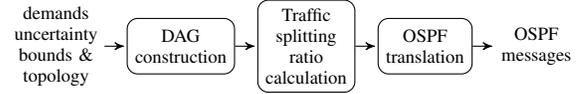
\begin{figure}
	  \centering
	\begin{small}
	\tikzset{
	>=stealth',
	  invis/.style={
	    draw=none,
	    text centered, 
	    align=center,
	    on chain},
	  punktchain/.style={
	    rectangle, 
	    rounded corners, 
	    align=center,
	    draw=black, thin,
	    inner sep=3pt,
	    minimum height=3em, 
	    text centered, 
	    on chain},
	  line/.style={draw, thin, <-},
	  every join/.style={->, thin,shorten >=1pt},
	}
	  \begin{tikzpicture}
	    [scale=.2,node distance=1.2em, start chain=going right,] %
	     \scriptsize 
	    \node[invis, join] (input) {demands\\uncertainty\\bounds \&\\topology}; %
	    \node[punktchain, join] (dag) [align=center]{DAG\\construction}; %
	    \node[punktchain, join] (gp) {Traffic\\splitting\\ratio\\calculation};%
	    \node[punktchain, join] (fibbing) {OSPF\\translation}; %
	    \node[invis, join] (output) {OSPF\\messages}; %
	  \end{tikzpicture}
	\end{small}%
	  \caption{COYOTE architecture.}%
	  \label{fig:COYOTE-architecture}
	\end{figure}

Figure~\ref{fig:COYOTE-architecture} presents an overview of the COYOTE architecture. COYOTE gets as input the (capacitated) network topology and the so-called ``uncertainty bounds'', i.e., for every two nodes (routers) in the network, $s$ and $t$, a real-valued interval $[d_{st}^{min},d_{st}^{max}]$, capturing the operator's uncertainty about the traffic demand from $s$ to $t$ or, alternatively, the potential variability of traffic. COYOTE then uses this information first to compute a forwarding DAG rooted in each destination node, and then to optimize traffic splitting ratios within each DAG. Lastly, the outcome of this computation is converted into OSPF configuration by injecting ``lies'' into routers. We next elaborate on each of these components.

\subsection{Computing DAGs}\label{sec:computing-dags} 

In COYOTE, DAGs rooted in different destinations are not coupled in any way, allowing network operators to specify any set of DAGs. We show, in Section~\ref{sec:eval}, however, that the following simple approach generates empirically good routing outcomes.

\vspace{0.1in}\noindent\textbf{Step I: Shortest-path DAG generation.} We assign each link a weight to generate a shortest-path DAG rooted in each destination (as in traditional OSPF routing). We evaluate in Section~\ref{sec:eval} two heuristics for setting link weights from the OSPF-ECMP TE literature:

\begin{itemize}

\item {\bf Reverse capacities.} Link weights are set to be the inverse of link capacities. We point out that this is compatible with Cisco's recommendations for default OSPF link weights~\cite{cisco-link-weights}.

\item {\bf Local search.} This heuristic leverages the techniques in~\cite{Altin:2012:OOR:2345370.2345375} for optimizing oblivious ECMP routing configurations. Specifically, link weights are initially set to be the reverse link capacities (as above). Then, the heuristic iteratively computes a worst-case traffic matrix for ECMP TE with respect to the current link weights, adds this matrix to a set of traffic matrices $\mathcal{T}$ (initially set to be empty), and myopically changes a single link's weight if this improves the worst-case \ecmp link utilization over the matrices in set $\mathcal{T}$. The reader is referred to Appendix~\ref{sec:local-search-dag} for a detailed exposition.
\end{itemize}

\noindent\textbf{Step II: DAG augmentation. } Once the shortest-path DAGs are computed, each DAG is \emph{augmented} with additional links as follows. Each link that does not appear in the shortest-path DAG for some target vertex $t$ is oriented towards the incident node that is closer to the destination $t$, breaking ties lexicographically (suppose that the nodes are numbered).

Let us revisit our running example in Fig.\ref{fig:sample-net}. Observe that while the shortest-path DAG rooted in $t$ does \emph{not} contain link $(s_2,v)$ if all links have the same weight, the augmented forwarding DAGs will also utilize this link (in some direction). DAG-augmentation allows us to enhance path diversity, and so increases the available network capacity. Since the final DAGs contain the original shortest-path DAGs, traditional ECMP routing is a point in the solution space over which COYOTE optimizes. COYOTE is thus guaranteed to compute an oblivious solution that is no worse than standard OSPF\slash ECMP. We show in Section~\ref{sec:eval} that COYOTE indeed significantly outperforms TE with ECMP using any of the two heuristics.

\subsection{In-DAG Traffic Splitting}\label{sec:splitting-ratio-overview}

Once per-destination DAGs are computed, as described above, COYOTE executes an algorithm that receives as input a set of per-destination DAGs and optimizes traffic splitting \emph{within} these DAGs, with the objective of minimizing the worst-case congestion (link utilization) over a given set of possible traffic demand matrices. 

Whether the problem of computing traffic splitting ratios within a set of given DAGs can be solved optimally in a computationally-efficient manner remains an open question (see Section~\ref{sec:conc}). This seems impossible within the familiar mathematical toolset of TE, namely, integer and linear programming. We found that a different approach is, however, feasible: casting the optimization problem described in Section~\ref{sec:notation-problem} as a geometric program (in fact, a mixed linear-geometric program (MLGP) \cite{boyd:geometric_prog}).
Stating COYOTE's traffic splitting optimization as a geometric program is not straightforward and involves careful application of various techniques (convex programming, monomial approximations, LP duality). We provide an intuitive exposition of some of these ideas below using the running example in Fig.~\ref{fig:COYOTE-overview}. We dive into the many technical details involved in computing COYOTE's traffic splitting ratios in Appendix~\ref{sec:dual-gp}.

Again, $s_1$ and $s_2$ send traffic to $t$, let the DAG for $t$ be as in Fig. \ref{fig:opt-COYOTE}, and suppose that the capacity on links $(s_1, s_2)$, $(s_1, v)$, and $(s_2, v)$ is infinite (that is, arbitrarily large) and on $(s_2, t)$ and $(v, t)$ is $1$.  We are given as input a set of possible demand matrices ${\cal D}$ for the two users and our goal is to find the traffic splitting ratios $\phi$ so that the worst-case link utilization across these two demand matrices is minimized. We assume, without loss of generality (since the performance ratio is invariant to rescaling) that traffic demands in ${\cal D}$ can always be routed without exceeding the link capacities. A simplified mathematical program for this problem would take the following form (see explanations below): %
\allowdisplaybreaks %
\begin{align}
  &\min \alpha \nonumber\\
  & \forall (d_{s_1t}, d_{s_2t}) \in {\cal D} \\
  &~~\frac{d_{s_1t}  \phi(s_1, s_2)  \phi(s_2, t) + d_{s_2t}  \phi(s_2, t)}{c_{s_2, t}} \le \alpha \label{eq:link_b}\\
  &~~\frac{d_{s_1t} (1 - \phi(s_1, s_2) \phi(s_2, t)) + d_{s_2t}\left(1-\phi(s_2, t)\right)}{c_{(v,t)}} \le \alpha \label{eq:link_c}
\end{align}
\vspace{-1.3em}%

The objective is to minimize $\alpha$, which represents worst-case link utilization, i.e., the load (flow divided by capacity) on the most utilized link across all the admissible demand matrices. Each variable $\phi(x,y)$ denotes the fraction of the incoming flow destined for to $t$ at vertex $x$ that is routed on link $(x,y)$. 
Constraints~\eqref{eq:link_b} and~\eqref{eq:link_c} force $\alpha$ to be at least the value of the link utilization of links $(s_2,t)$ and $(s_1,s_2)$, respectively. Since the other links have infinite (arbitrarily high) capacities, the load on these links is negligible and so the link utilization inequalities for these links are omitted. Now, consider Constraint~\eqref{eq:link_b} for link $(s_2,t)$. Observe that from user $s_1$ the fraction of traffic sent through $(s_2,t)$ equals the fraction of $s_1$'s traffic through $(s_1,s_2)$ (i.e., $\phi(s_1,s_2)$) times the fraction sent through $(s_2,t)$ by $s_2$ (i.e., $\phi(s_2,t)$). The fraction of $s_2$'s traffic through $(s_2,t)$ is simply $\phi(s_2,t)$. Accordingly the total flow on $(s_2,t)$ equals $d_{s_1t} \cdot \phi(s_1,s_2) \cdot \phi(s_2,t) + d_{s_2t} \cdot \phi(s_2,t)$. Hence, the link utilization of $(s_2,t)$ is this expression divided by the capacity of $(s_2,t)$, and the corresponding constraint \eqref{eq:link_b} requires that this utilization be at most $\alpha$ for \emph{all} demand matrices $(d_{s_1}, d_{s_2}) \in {\cal D}$. Constraint \eqref{eq:link_c} states the same for link $(v,t)$, where the fraction of traffic sent by $s_1$ ($s_2$) to $t$ through $(v,t)$ is equal to $1$ minus the fraction of flow sent from $s_1$ ($s_2$) to $t$ through $(s_2,t)$.

Two difficulties with these constraints immediately arise: one is that it is \emph{universally quantified} over an entire set of  demand matrices, possibly of infinite cardinality, and the other is that it involves a \emph{product of unknowns}, namely, $\phi(s_1,s_2) \cdot \phi(s_2,t)$, and such products do not fit into the framework of standard linear and integer programming. For a discrete set of demand matrices we can handle the first problem by stating \eqref{eq:link_b} and \eqref{eq:link_c} for each individual demand matrix. Otherwise (if the set of DMs  is of infinite size) the elegant dualization technique from~\cite{4032716}, which we describe in Appendix~\ref{sec:dual-gp}, can be used.  To handle the second issue, however, we need a small trick from geometric programming~\cite{boyd:geometric_prog}.  Let $d_{s_1t}=1$ and $d_{s_2t}=1$ and consider constraint \eqref{eq:link_b}:
\begin{displaymath}
  \phi(s_2,t) + \phi(s_1,s_2) \cdot \phi(s_2,t) \le \alpha \enspace .
\end{displaymath}

Now, let $\widetilde{\phi}(s_1,s_2) = \log \phi(s_1,s_2)$ and $\widetilde{\phi}(s_2,t) = \log \phi(s_2,t)$, and take the logarithm of both sides:
\begin{displaymath}
  \log\left(e^{\widetilde{\phi}(s_2,t)} + e^{\widetilde{\phi}(s_1,s_2) + \widetilde{\phi}(s_2,t)} \right)\le \log \alpha \enspace .
\end{displaymath}

This constraint is now a logarithm of a sum of exponentials of linear functions and so is convex, opening the door to using standard convex programming. 
Our implementation uses a convex program based on the above ideas (and other ingredients) to compute the traffic splitting ratios. We provide a detailed exposition in Appendix~\ref{sec:dual-gp}.

\subsection{Translation to OSPF-ECMP configuration.} 

As explained above, using OSPF and ECMP for TE constrains the flow of traffic in two significant ways: (1) traffic only flows on shortest-paths (induced from operator specified link weights), and (2) traffic is split equally between multiple next-hops on shortest-paths to a destination. Recent studies show how OSPF-ECMP's expressiveness can be significantly enhanced by effectively deceiving routers. Specifically, Fibbing~\cite{Vissicchio:2014:SLL:2670518.2673868,Vissicchio:2015:CCO:2785956.2787497} shows how \emph{any} set of per-destination forwarding DAGs can be realized by introducing fake nodes and virtual links into an underlying link-state routing protocol, thus overcoming the first limitation of ECMP. \cite{nemeth:networking_2013} shows how ECMP's equal load balancing can be extended to much more nuanced traffic splitting by setting up virtual links alongside existing physical ones, thus relaxing the second of these limitations.

We revisit our running example to show how COYOTE exploits these techniques. Consider Fig.~\ref{fig:COYOTE-fibbing}. Inserting a fake advertisement at $s_1$ into the OSPF link-state database can ``deceive'' $s_1$ into believing that, besides its available shortest paths via $s_2$ and $v$ , destination $t$ is also available via a third, ``virtual'' forwarding path. The forwarding adjacency in the fake advertisement is mapped to $s_2$, so that out of $s_1$'s three next-hops to $t$ node $s_2$ will appear \emph{twice} while $v$ only appears once. Consequently, the traffic is \emph{effectively} split between $s_2$ and $v$ in a ratio \nicefrac{2}{3} to \nicefrac{1}{3}. Beyond changing how traffic is split \emph{within} a given shortest-path DAG, as illustrated in Fig.~\ref{fig:COYOTE-fibbing}, fake nodes\slash links can be injected into OSPF so to as change the forwarding DAGs themselves at the per-IP-destination-prefix granularity, as shown in~\cite{Vissicchio:2015:CCO:2785956.2787497}. COYOTE leverages the techniques in~\cite{Vissicchio:2015:CCO:2785956.2787497} and in~\cite{nemeth:networking_2013} to carefully craft ``lies'' so as to generate the desired per-destination forwarding DAGs and approximate the optimal traffic splitting ratios with ECMP. We show in Section~\ref{sec:eval} that highly optimized TE is achievable even with the introduction of few virtual nodes and links.



%% file: 600-tr-evaluation.tex
\section{Evaluation}
\label{sec:eval}
\begin{figure*}[!ht]
	\centering
	\begin{minipage}{\linewidth}
		\centering
		\frame{\includegraphics[width=.9\columnwidth]{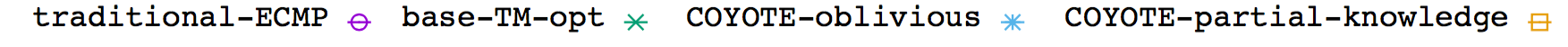}}
		\caption*{}
		\label{fig:leged}
	\end{minipage}
	
	\vspace{-.2in}
	
	\begin{minipage}{.32\linewidth}
		\centering
		\includegraphics[width=\columnwidth]{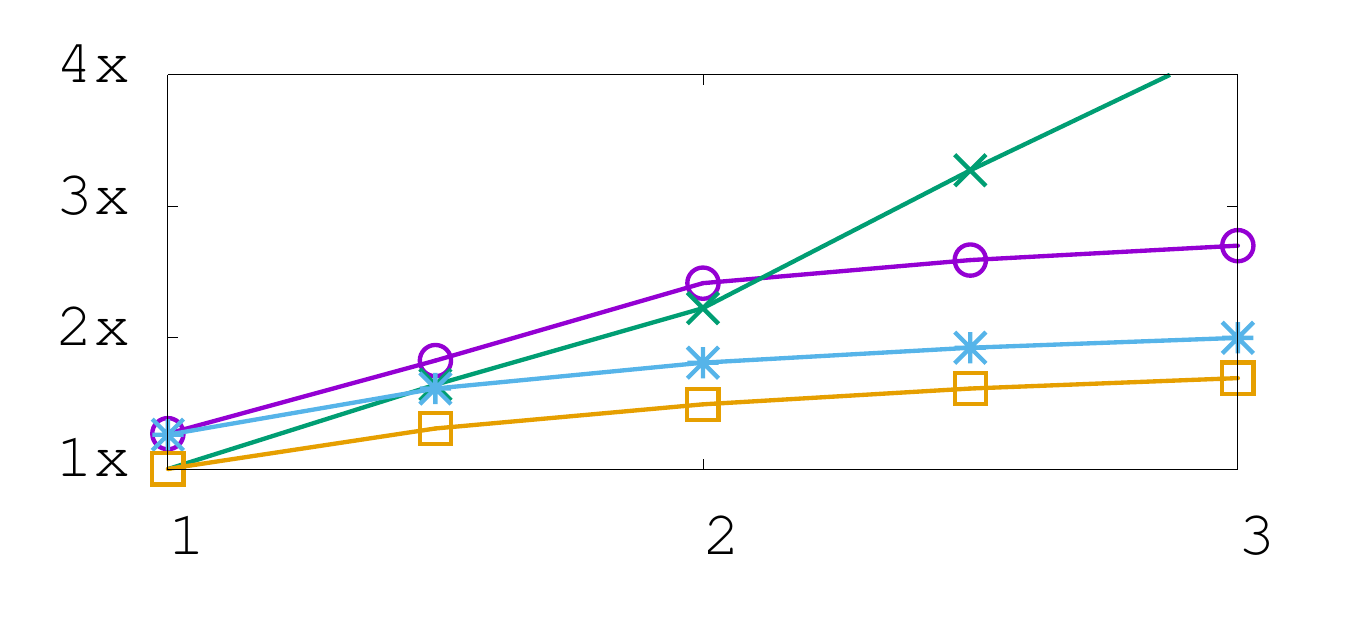}
		\caption{Geant, gravity model.}
		\label{fig:gravity-augmented-shortest-path-dag-geant}
	\end{minipage}
	\begin{minipage}{.32\linewidth}
		\centering
		\includegraphics[width=\columnwidth]{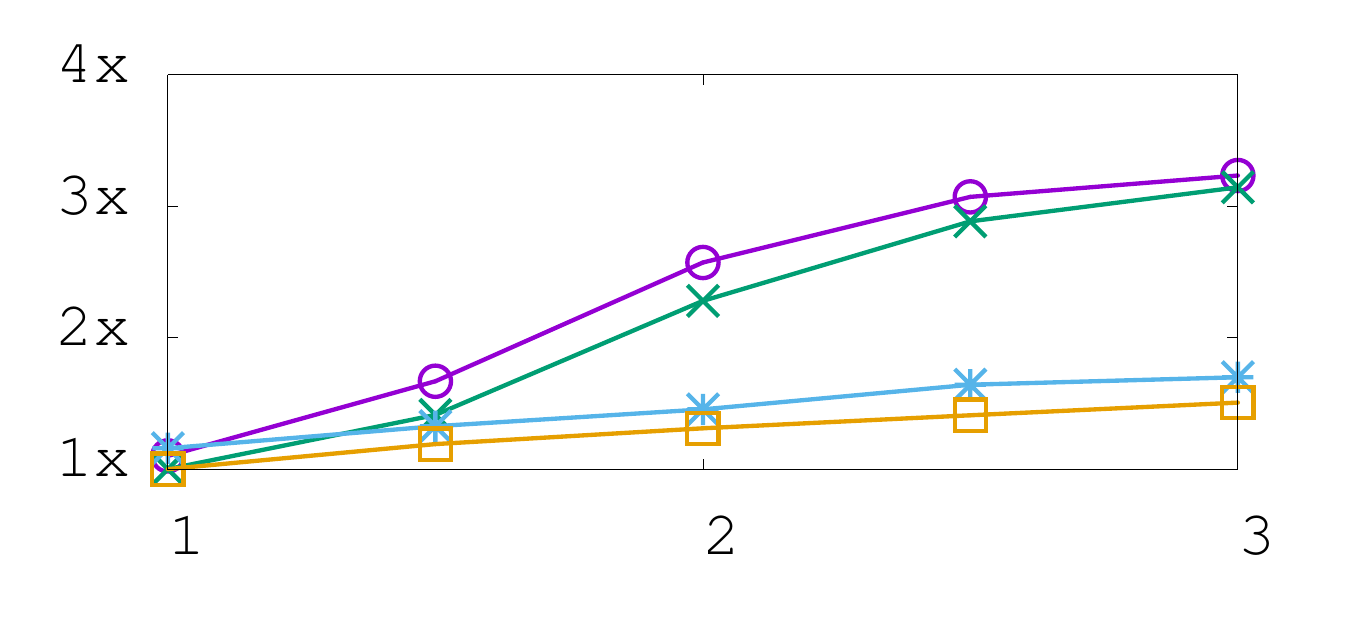}
		\caption{Digex, gravity model.}
		\label{fig:gravity-augmented-shortest-path-dag-digex}
	\end{minipage}
	\begin{minipage}{.32\linewidth}
		\centering
		\includegraphics[width=\columnwidth]{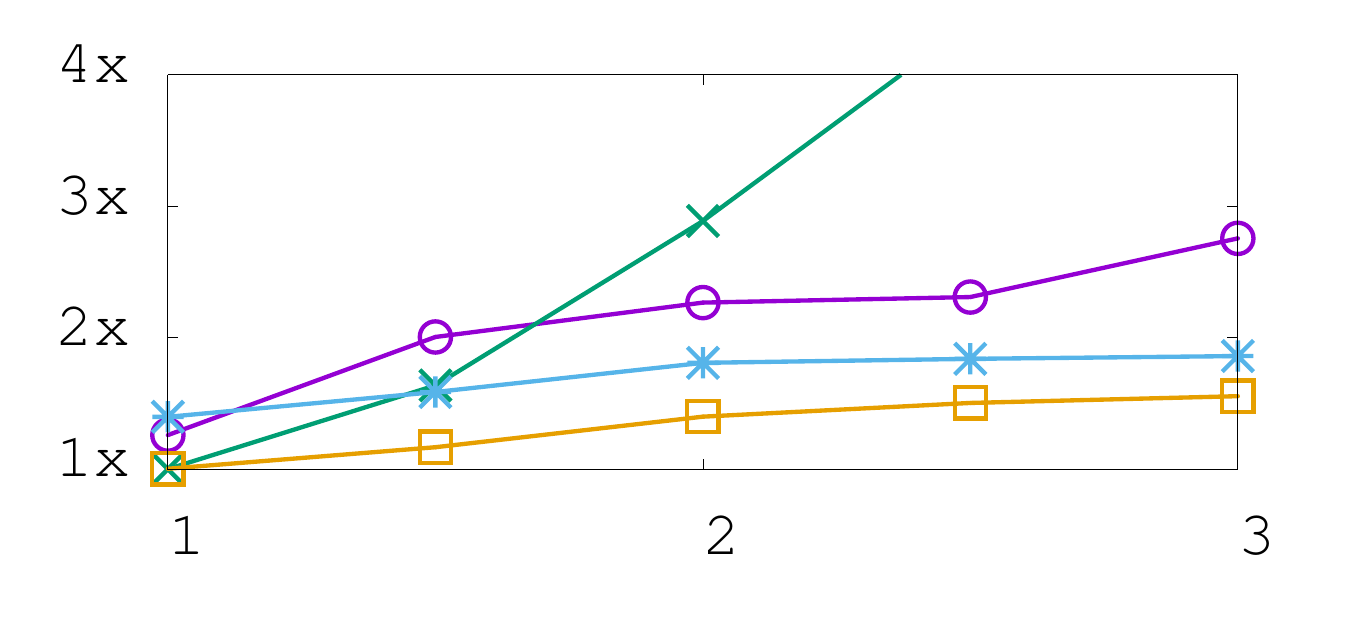}
		\caption{AS 1755, bimodal model}
		\label{fig:bimodal-augmented-shortest-path-dag}
	\end{minipage}
	
	\centering
	
	\begin{minipage}{.32\linewidth}
		\centering
		\includegraphics[width=\columnwidth]{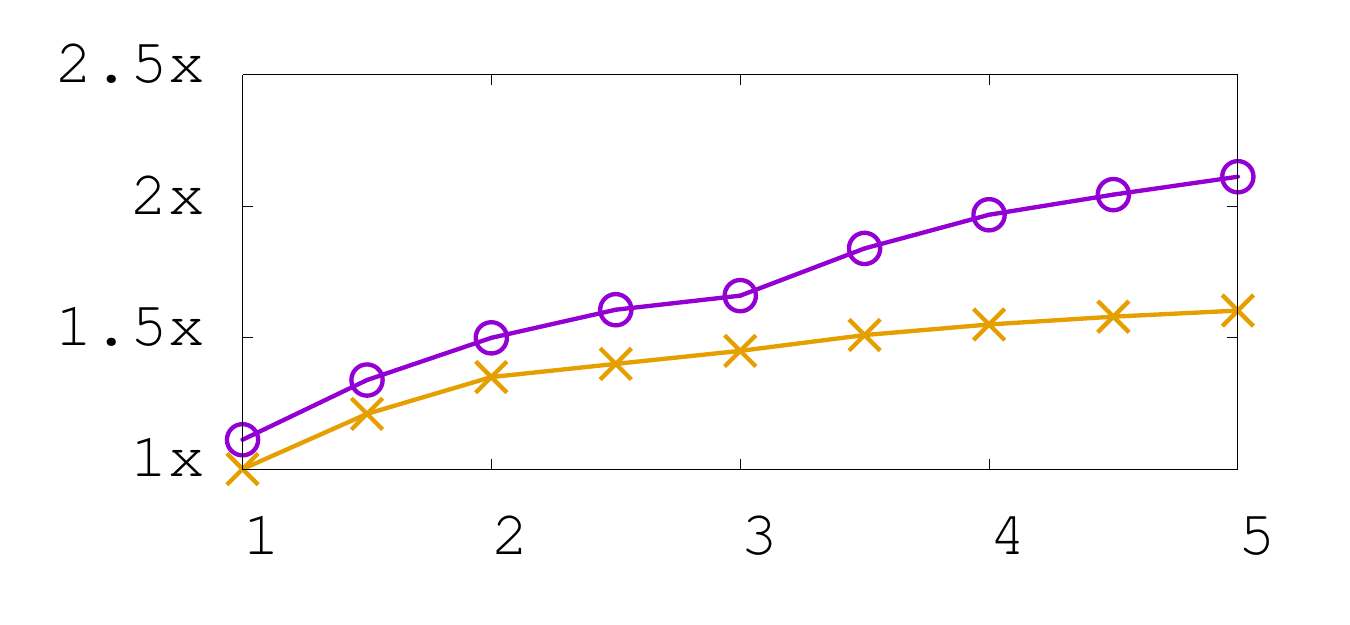}
		\caption{Abilene, local search heuristic.}   
		\label{fig:obliviuos-ecmp}
	\end{minipage}
	\begin{minipage}{.32\linewidth}
		\centering
		\includegraphics[width=\columnwidth]{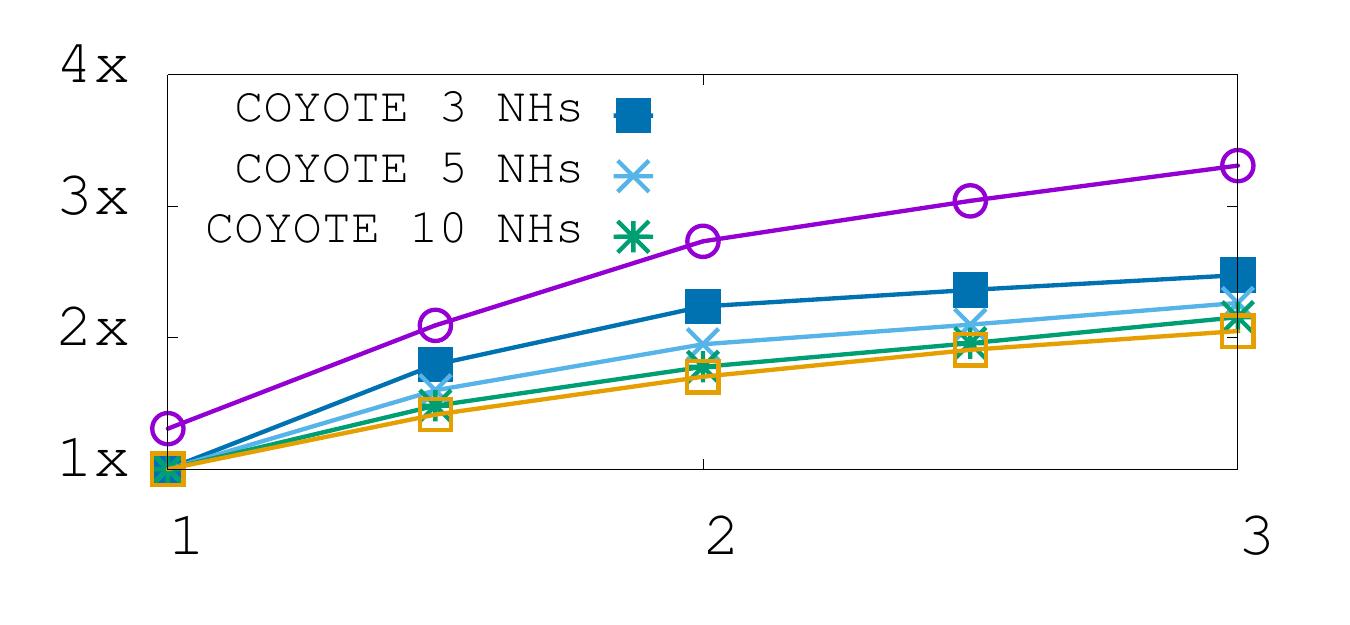}
		\caption{Approximation.}
		\label{fig:approximation}
	\end{minipage}
	\begin{minipage}{.32\linewidth}
		\centering
		\includegraphics[width=\columnwidth]{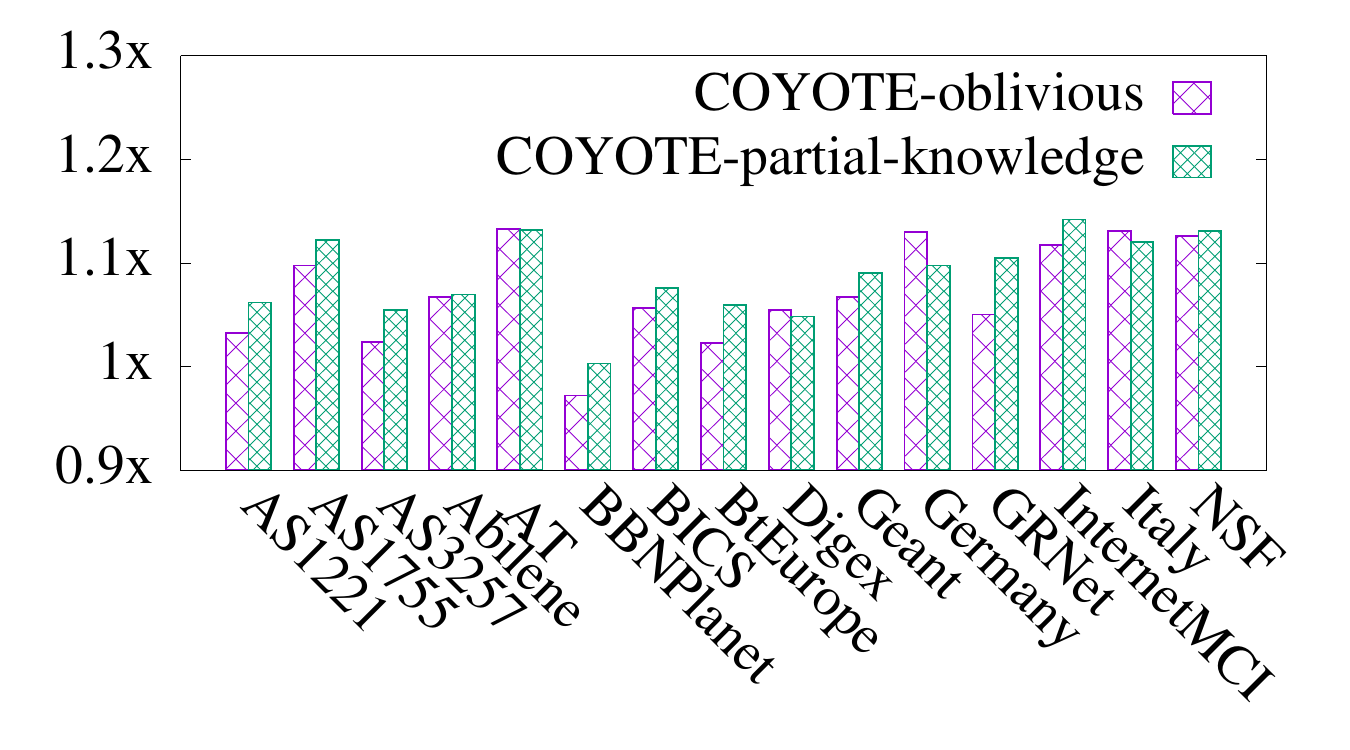}
		\caption{Average stretch.}
		\label{fig:stretch}
	\end{minipage}

\end{figure*}

We experimentally evaluate COYOTE in order to quantify its performance benefits and its robustness to traffic uncertainty and variation. Importantly, our focus is solely on destination-based TE schemes (i.e., TE schemes that can be realized via today's IP routing). We show below that COYOTE provides significantly better performance than ECMP even when \emph{completely} oblivious to the traffic demand matrices. Also, COYOTE's increased path diversity does not come at the cost of long paths: the paths computed by COYOTE are on average only a factor of 1.1 longer than \ecmp's. We also discuss experiments with a prototype implementation of COYOTE.

While the reader might think that COYOTE's performance benefits over traditional TE with ECMP are merely a byproduct of its greater flexibility in selecting DAGs and in traffic splitting, our results show that this intuition is, in fact, false. Specifically, we show that, similarly to unconstrained (i.e., source and destination based) oblivious routing~\cite{4032716}, even the \emph{optimal} routing with respect to \emph{estimated}  demand matrices fares much worse than COYOTE if the \emph{actual} demand matrices are not very ``close'' to the estimated demands. Hence, COYOTE's good performance should be attributed not only to its expressiveness but also, in large part, to its built-in algorithms for optimizing performance in the presence of uncertainty, as discussed in Section~\ref{sec:design}.

\subsection{Simulation Framework} 

We use the set of $16$ backbone Internet topologies from the Internet Topology Zoo (ITZ) archive~\cite{topology-zoo} to assess the performance of COYOTE and \ecmp. When available, we use the link capacities provided by ITZ. Otherwise, we set the link capacities to be inversely-proportional to the ITZ-provided \ecmp weights (in accordance with the Cisco-recom\-mend\-ed default OSPF link configuration~\cite{cisco-link-weights}). When neither \ecmp link weights nor capacities are available we use unit capacities and link weights.
We evaluate COYOTE against ECMP using the two simple DAG-construction heuristics described in Section~\ref{sec:computing-dags} : reverse capacities and local search.

To compute COYOTE's in-DAG traffic splitting ratios (see Section~\ref{sec:design}), we use AMPL~\cite{ampl} as the problem formulation language and MOSEK~\cite{mosek}, a non-linear convex optimization solver. The running time with our current single-threaded proof-of-concept implementation ranges from few minutes (for small networks) to few days (for large networks). 

We would like to point out that the computation of the in-DAG traffic splitting ratios needs only be performed
once or on a daily/weekly-base, as routing in COYOTE is not dynamically adjusted, and that routing configurations for failure scenarios (e.g., every single link/node failure) can be precomputed.

We measure performance in terms of the worst-case link utilization (referred to as ``congestion'' in TE literature~\cite{1039866, 4483669}), i.e., the performance of (multicommodity) flow of traffic $f$ is $\max_{l}\frac{f(l)}{c_l}$, where $f(l)$ is the flow traversing link $l$ and $c_l$ is link $l$'s capacity.

\subsection{Network Performance}

We compare COYOTE to ECMP for both DAG-construction heuristics described above and for two types of base  demand matrices: (1) \emph{gravity}~\cite{Roughan:2002:EMB:637201.637213}, where the amount of flow sent from router $i$ to router $j$ is proportional to the product of $i$'s and $j$'s total outgoing capacities, and (2) \emph{bimodal}~\cite{Medina:2002:TME:964725.633041}, where a small fraction of all pairs of routers exchange large quantities of traffic, and the other pairs send small flows.

We first present our results with respect to the reverse capacities heuristic, which is  based on ITZ~\cite{topology-zoo} link weights, and an ideal version of COYOTE capable of arbitrarily fine-grained traffic splitting. We then show that a close approximation of the optimal splitting ratios can be obtained with the introduction of a limited number of additional virtual links.
Fig.~\ref{fig:gravity-augmented-shortest-path-dag-geant} and Fig.~\ref{fig:gravity-augmented-shortest-path-dag-digex} describe our results for two networks (Geant and Digex, respectively), the gravity model, and augmented shortest path DAGs based on the ITZ link weights. The x-axis represents the ``uncertainty margin'': let $d_{st}$ be the amount of flow from router $s$ to router $t$ in the base  demand matrices (namely, gravity), a margin of uncertainty of $x$ means that the actual flow from $s$ to $t$ can be any value between $\frac{d_{st}}{x}$ and $x\cdot d_{st}$. We increase the uncertainty margin in increments of $0.5$ from $1$ (no uncertainty whatsoever) to $3$ (fairly high uncertainty). The y-axis specifies how far the computed solution is from the \emph{demands-aware} optimum within the same DAGs. 

We plot four lines, corresponding to the performance of four different protocols: (1) traditional TE with ECMP, (2) the optimal demands-aware routing for the base gravity model (with no uncertainty), which can be obtained with linear programming techniques~\cite{multicommodityflow-lp}, (3) COYOTE (oblivious) with traffic splitting optimized with respect to \emph{all} possible  demand matrices (i.e., assuming \emph{nothing} about the demands), (4) COYOTE (partial-knowledge) optimized with respect to the  demand matrices within the uncertainty margin. Observe that both variants of COYOTE provide significantly better performance than TE with ECMP and, more surprisingly, both COYOTE and (sometimes) ECMP outperform the optimal base routing, whose performance quickly degrades even with little demands uncertainty. Our results thus show that COYOTE's built-in robustness to traffic uncertainty, in the form of optimization under specified uncertainty margins, indeed leads to superior performance in the face of inaccurate knowledge about the demand matrices or, alternatively, variable traffic conditions.
Table~\ref{tab:evaluation} shows  the extensive results of COYOTE for all the analyzed topologies, except BBNPlanet and Gambia, which are almost a tree topology.
\input{950-appendix-evaluation}

We observe the same trends when the base demand matrices are sampled from the bimodal model, as shown in Fig.~\ref{fig:bimodal-augmented-shortest-path-dag}.

We now discuss our results for the local search DAG-construction heuristic (see Section~\ref{sec:computing-dags}).
Fig.~\ref{fig:obliviuos-ecmp} presents a comparison of COYOTE and \ecmp using the bimodal model as the base demand matrices. We use the above heuristic to compute, for each uncertainty margin in the range $1-5$, increasing in $0.5$ increments, the (traditional) ECMP configuration and COYOTE DAGs with respect to the bimodal-based demand matrices. We then compare the worst-case link utilization of the two, again, normalized by the \emph{demands-aware} optimum within the same (augmented) DAGs. We note that \ecmp is, on average, almost 80\% times further away from the optimum than COYOTE.

\vspace{0.05in}\noindent\textbf{Approximating the optimal traffic splitting.} We evaluated above COYOTE under the assumption that arbitrarily fine-grained traffic splitting is achievable, yet in practice, the resolution of traffic splitting is derived from the number of virtual links introduced. Clearly, an excessive number of virtual links should be avoided for at least two reasons: (a) each virtual next-hop is installed into the finite-sized Forwarding Information Base (FIB), and (b) injecting additional information into OSPF comes at the cost of additional computational overhead. Our results, illustrated in Fig.~\ref{fig:approximation} for AS 1755 network's topology (all other topologies exhibit the same trend), show that even with just $3$ additional virtual links per router interface, COYOTE achieves a 50\% improvement over traditional TE with ECMP. We observe that with $10$ virtual links the computed routing configuration closely approximates the ideal solution.

\vspace{0.05in}\noindent\textbf{Average path lengths.}
COYOTE augments the shortest path DAG with additional links so as to better utilize the network. Consequently, traffic can potentially traverse longer paths. We show, however, that COYOTE's increased path redundancy does not come at the expense of long paths. Specifically, the average stretch (increase in length) of the paths in COYOTE is typically bounded within a 10\% factor with respect to the OSPF/ECMP paths. Fig.~\ref{fig:stretch} plots the average stretch across all pairs for a margin of $2.5$. Similar results are obtained for all different margins between $1$ to $5$. Observe that the DAGs computed by COYOTE rely on shortest-path computation with respect to the link weights, whereas the stretch is measured in terms of the number of hops. Thus, it is possible for the stretch to be less than 1, as is the case, e.g., for BBNPlanet.

%% file: 950-appendix-evaluation.tex
\begin{table*}
	\small
	\def\arraystretch{0.95}
	\centering
	\caption{Comparison of COYOTE against traditional ECMP and Base-TM-opt for the gravity base model.}
	\begin{minipage}{.45\linewidth}
		\begin{tabular}{cc|c|c|cc}
			\toprule
			& &  &  &  \multicolumn{2}{c}{{\bf COYOTE} }   \\
			\bf Network & \bf margin & {\bf ECMP } & {\bf Base}& {\bf obl.} &  {\bf par.know.}\\
			\midrule
			\texttt{\multirow{9}{*}{\rotatebox[origin=c]{90}{
						\begin{minipage}{1.4cm}
							\centering
							1221 
						\end{minipage}
					}}}
					& 1.0 & 1.00 & 1.00 & 1.00 & 1.00 \\
					& 1.5 & 1.00 & 1.00 & 1.00 & 1.00 \\
					& 2.0 & 1.00 & 1.00 & 1.00 & 1.00 \\
					& 2.5 & 1.00 & 1.15 & 1.00 & 1.00 \\
					& 3.0 & 1.00 & 1.34 & 1.00 & 1.00 \\
					& 3.5 & 1.17 & 1.65 & 1.14 & 1.00 \\
					& 4.0 & 1.29 & 2.09 & 1.30 & 1.07 \\
					& 4.5 & 1.36 & 2.59 & 1.49 & 1.20 \\
					& 5.0 & 1.68 & 3.15 & 1.52 & 1.30 \\
					\midrule
					\texttt{\multirow{9}{*}{\rotatebox[origin=c]{90}{
								\begin{minipage}{1.4cm}
									\centering
									1755 
								\end{minipage}
							}}}
							& 1.0 & 1.31 & 1.00 & 1.27 & 1.00 \\
							& 1.5 & 2.09 & 2.15 & 2.04 & 1.41 \\
							& 2.0 & 2.73 & 3.72 & 2.19 & 1.70 \\
							& 2.5 & 3.04 & 5.79 & 2.32 & 1.91 \\
							& 3.0 & 3.31 & 7.71 & 2.39 & 2.05 \\
							& 3.5 & 3.64 & 9.26 & 2.47 & 2.16 \\
							& 4.0 & 3.90 & 10.65 & 2.51 & 2.24 \\
							& 4.5 & 4.04 & 11.85 & 2.55 & 2.30 \\
							& 5.0 & 4.13 & 12.83 & 2.57 & 2.36 \\
							\midrule
							\texttt{\multirow{9}{*}{\rotatebox[origin=c]{90}{
										\begin{minipage}{1.4cm}
											\centering
											3257 
										\end{minipage}
									}}}
									& 1.0 & 5.06 & 1.00 & 1.63 & 1.00 \\
									& 1.5 & 6.56 & 2.25 & 2.30 & 1.55 \\
									& 2.0 & 7.53 & 3.85 & 2.69 & 1.95 \\
									& 2.5 & 8.18 & 5.86 & 2.91 & 2.22 \\
									& 3.0 & 8.62 & 7.27 & 2.98 & 2.83 \\
									& 3.5 & 8.87 & 9.20 & 3.02 & 3.60 \\
									& 4.0 & 8.99 & 11.15 & 3.05 & 3.04 \\
									& 4.5 & 9.07 & 12.99 & 3.23 & 3.15 \\
									& 5.0 & 9.14 & 14.76 & 3.47 & 2.82 \\
									\midrule
									\texttt{\multirow{9}{*}{\rotatebox[origin=c]{90}{
												\begin{minipage}{1.4cm}
													\centering
													BICS 
												\end{minipage}
											}}}
											& 1.0 & 2.62 & 1.00 & 1.73 & 1.00 \\
											& 1.5 & 2.82 & 2.01 & 1.95 & 1.21 \\
											& 2.0 & 2.90 & 3.21 & 2.00 & 1.26 \\
											& 2.5 & 2.93 & 4.86 & 2.02 & 1.36 \\
											& 3.0 & 2.95 & 6.89 & 2.03 & 1.51 \\
											& 3.5 & 3.02 & 9.28 & 2.04 & 1.62 \\
											& 4.0 & 3.13 & 11.87 & 2.04 & 1.69 \\
											& 4.5 & 3.21 & 14.66 & 2.05 & 1.76 \\
											& 5.0 & 3.27 & 17.77 & 2.07 & 1.82 \\
											\midrule
											\texttt{\multirow{9}{*}{\rotatebox[origin=c]{90}{
														\begin{minipage}{1.4cm}
															\centering
															BtEurope 
														\end{minipage}
													}}}
													& 1.0 & 1.16 & 1.00 & 1.15 & 1.00 \\
													& 1.5 & 1.17 & 2.15 & 1.16 & 1.08 \\
													& 2.0 & 1.33 & 3.60 & 1.22 & 1.11 \\
													& 2.5 & 1.73 & 5.24 & 1.55 & 1.12 \\
													& 3.0 & 1.91 & 6.95 & 1.61 & 1.14 \\
													& 3.5 & 1.95 & 8.66 & 1.64 & 1.20 \\
													& 4.0 & 2.06 & 10.31 & 1.77 & 1.27 \\
													& 4.5 & 2.53 & 11.93 & 2.14 & 1.35 \\
													& 5.0 & 2.95 & 14.47 & 2.31 & 1.34 \\
													\midrule
													\texttt{\multirow{9}{*}{\rotatebox[origin=c]{90}{
																\begin{minipage}{1.4cm}
																	\centering
																	Digex 
																\end{minipage}
															}}}
															& 1.0 & 1.10 & 1.00 & 1.16 & 1.00 \\
															& 1.5 & 1.67 & 1.41 & 1.33 & 1.19 \\
															& 2.0 & 2.57 & 2.28 & 1.45 & 1.31 \\
															& 2.5 & 3.07 & 2.89 & 1.64 & 1.41 \\
															& 3.0 & 3.23 & 3.14 & 1.70 & 1.51 \\
															& 3.5 & 3.33 & 3.26 & 1.77 & 1.57 \\
															& 4.0 & 3.43 & 3.34 & 1.89 & 1.64 \\
															& 4.5 & 3.51 & 3.42 & 1.99 & 1.72 \\
															& 5.0 & 3.58 & 3.49 & 2.06 & 1.78 \\
															\midrule
															\texttt{\multirow{9}{*}{\rotatebox[origin=c]{90}{
																		\begin{minipage}{1.4cm}
																			\centering
																			GRNet 
																		\end{minipage}
																	}}}
																	& 1.0 & 1.88 & 1.00 & 1.32 & 1.00 \\
																	& 1.5 & 1.97 & 1.70 & 1.53 & 1.32 \\
																	& 2.0 & 2.10 & 2.26 & 1.65 & 1.44 \\
																	& 2.5 & 2.14 & 2.70 & 1.84 & 1.51 \\
																	& 3.0 & 2.20 & 3.07 & 2.15 & 1.59 \\
																	& 3.5 & 2.52 & 3.34 & 2.41 & 1.65 \\
																	& 4.0 & 2.78 & 3.56 & 2.65 & 1.70 \\
																	& 4.5 & 2.96 & 3.72 & 2.86 & 1.73 \\
																	& 5.0 & 3.11 & 3.84 & 3.04 & 1.75 \\
																	\bottomrule 
																\end{tabular}
																
															\end{minipage}
															\begin{minipage}{.45\linewidth}
																\begin{tabular}{cc|c|c|cc}
																	\toprule
																	& &  &  &  \multicolumn{2}{c}{{\bf COYOTE} }   \\
																	\bf Network & \bf margin & {\bf ECMP } & {\bf Base}& {\bf obl.} &  {\bf par.know.}\\
																	\midrule
																	\texttt{\multirow{9}{*}{\rotatebox[origin=c]{90}{
																				\begin{minipage}{1.4cm}
																					\centering
																					Geant 
																				\end{minipage}
																			}}}
																			& 1.0 & 1.27 & 1.00 & 1.26 & 1.00 \\
																			& 1.5 & 1.83 & 1.64 & 1.61 & 1.31 \\
																			& 2.0 & 2.42 & 2.22 & 1.81 & 1.49 \\
																			& 2.5 & 2.59 & 3.27 & 1.92 & 1.61 \\
																			& 3.0 & 2.70 & 4.25 & 2.00 & 1.69 \\
																			& 3.5 & 2.77 & 5.18 & 2.17 & 1.76 \\
																			& 4.0 & 2.82 & 6.04 & 2.29 & 1.81 \\
																			& 4.5 & 2.86 & 6.65 & 2.36 & 1.85 \\
																			& 5.0 & 2.88 & 6.75 & 2.41 & 1.88 \\
																			\midrule
																			\texttt{\multirow{9}{*}{\rotatebox[origin=c]{90}{
																						\begin{minipage}{1.4cm}
																							\centering
																							Germany cost 
																						\end{minipage}
																					}}}
																					& 1.0 & 1.42 & 1.00 & 1.23 & 1.00 \\
																					& 1.5 & 1.91 & 1.85 & 1.73 & 1.37 \\
																					& 2.0 & 2.23 & 2.45 & 2.03 & 1.68 \\
																					& 2.5 & 2.44 & 2.79 & 2.15 & 1.84 \\
																					& 3.0 & 2.58 & 3.03 & 2.23 & 1.95 \\
																					& 3.5 & 2.68 & 3.24 & 2.29 & 2.03 \\
																					& 4.0 & 2.75 & 3.39 & 2.33 & 2.09 \\
																					& 4.5 & 2.80 & 3.50 & 2.38 & 2.14 \\
																					& 5.0 & 2.84 & 3.59 & 2.41 & 2.19 \\
																					\midrule
																					\texttt{\multirow{9}{*}{\rotatebox[origin=c]{90}{
																								\begin{minipage}{1.4cm}
																									\centering
																									Internetmci 
																								\end{minipage}
																							}}}
																							& 1.0 & 1.07 & 1.00 & 1.30 & 1.00 \\
																							& 1.5 & 1.50 & 1.84 & 1.52 & 1.23 \\
																							& 2.0 & 2.49 & 3.22 & 2.04 & 1.67 \\
																							& 2.5 & 2.73 & 4.22 & 2.26 & 1.97 \\
																							& 3.0 & 2.95 & 4.61 & 2.37 & 2.12 \\
																							& 3.5 & 3.21 & 4.90 & 2.43 & 2.23 \\
																							& 4.0 & 3.39 & 5.13 & 2.49 & 2.31 \\
																							& 4.5 & 3.54 & 5.30 & 2.53 & 2.38 \\
																							& 5.0 & 3.66 & 5.49 & 2.56 & 2.43 \\
																							\midrule
																							\texttt{\multirow{9}{*}{\rotatebox[origin=c]{90}{
																										\begin{minipage}{1.4cm}
																											\centering
																											Italy cost 
																										\end{minipage}
																									}}}
																									& 1.0 & 1.70 & 1.00 & 1.31 & 1.00 \\
																									& 1.5 & 2.27 & 1.83 & 1.64 & 1.42 \\
																									& 2.0 & 2.57 & 2.41 & 2.08 & 1.68 \\
																									& 2.5 & 3.10 & 3.11 & 2.38 & 1.88 \\
																									& 3.0 & 3.39 & 3.89 & 2.56 & 2.01 \\
																									& 3.5 & 3.66 & 4.45 & 2.67 & 2.15 \\
																									& 4.0 & 3.91 & 4.91 & 2.75 & 2.28 \\
																									& 4.5 & 4.13 & 5.31 & 2.80 & 2.39 \\
																									& 5.0 & 4.35 & 5.54 & 2.85 & 2.48 \\
																									\midrule
																									\texttt{\multirow{9}{*}{\rotatebox[origin=c]{90}{
																												\begin{minipage}{1.4cm}
																													\centering
																													NSF cost 
																												\end{minipage}
																											}}}
																											& 1.0 & 1.71 & 1.00 & 1.51 & 1.00 \\
																											& 1.5 & 2.31 & 1.83 & 1.95 & 1.36 \\
																											& 2.0 & 2.71 & 2.49 & 2.20 & 1.60 \\
																											& 2.5 & 3.01 & 3.03 & 2.36 & 1.76 \\
																											& 3.0 & 3.25 & 3.32 & 2.47 & 1.87 \\
																											& 3.5 & 3.44 & 3.51 & 2.55 & 1.96 \\
																											& 4.0 & 3.62 & 3.70 & 2.61 & 2.06 \\
																											& 4.5 & 3.78 & 3.87 & 2.66 & 2.17 \\
																											& 5.0 & 3.93 & 4.00 & 2.70 & 2.25 \\
																											\midrule
																											\texttt{\multirow{9}{*}{\rotatebox[origin=c]{90}{
																														\begin{minipage}{1.4cm}
																															\centering
																															abilene cost 
																														\end{minipage}
																													}}}
																													& 1.0 & 1.15 & 1.00 & 1.06 & 1.00 \\
																													& 1.5 & 1.54 & 1.43 & 1.39 & 1.28 \\
																													& 2.0 & 1.71 & 1.63 & 1.53 & 1.42 \\
																													& 2.5 & 1.79 & 1.85 & 1.59 & 1.50 \\
																													& 3.0 & 2.02 & 2.33 & 1.83 & 1.57 \\
																													& 3.5 & 2.32 & 2.66 & 1.94 & 1.69 \\
																													& 4.0 & 2.49 & 2.88 & 2.00 & 1.79 \\
																													& 4.5 & 2.59 & 3.05 & 2.05 & 1.87 \\
																													& 5.0 & 2.66 & 3.22 & 2.08 & 1.92 \\
																													\midrule
																													\texttt{\multirow{9}{*}{\rotatebox[origin=c]{90}{
																																\begin{minipage}{1.4cm}
																																	\centering
																																	atnt cost 
																																\end{minipage}
																															}}}
																															& 1.0 & 2.22 & 1.00 & 1.56 & 1.00 \\
																															& 1.5 & 3.07 & 1.96 & 2.18 & 1.52 \\
																															& 2.0 & 3.76 & 2.89 & 2.39 & 1.88 \\
																															& 2.5 & 4.26 & 3.76 & 2.59 & 2.19 \\
																															& 3.0 & 4.60 & 4.41 & 2.75 & 2.41 \\
																															& 3.5 & 4.83 & 4.79 & 2.87 & 2.58 \\
																															& 4.0 & 5.01 & 5.07 & 2.95 & 2.70 \\
																															& 4.5 & 5.17 & 5.33 & 3.03 & 2.79 \\
																															& 5.0 & 5.28 & 5.64 & 3.08 & 2.87 \\
																															\bottomrule 
																														\end{tabular}
																														
																													\end{minipage}
																													\label{tab:evaluation}
																												\end{table*}

%% file: 700-tr-prototype.tex
\section{Prototype Implementation}\label{sec:prototype}
We implemented and experimented with a prototype of the COYOTE architecture, as described in Section~\ref{sec:design}. Our prototype extends the Fibbing controller code, written in Python and provided by Vissicchio et al.~\cite{Vissicchio:2015:CCO:2785956.2787497}, and uses the code of Nemeth et al. from~\cite{nemeth:networking_2013} for approximating the splitting ratios. 
We plan to make our code public in the near future. 
We next illustrate the benefits of COYOTE over traditional TE, as reflected by an evaluation of our prototype via the mininet~\cite{mininet-hifi} network emulator. 

\begin{figure}
	\centering
	\subfloat[][]{%
		\includegraphics[width=45pt,valign=c]{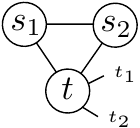}%
		\label{fig:mininet-example}}%
	\hfill%
	\subfloat[][]{%
		\includegraphics[width=180pt,valign=c]{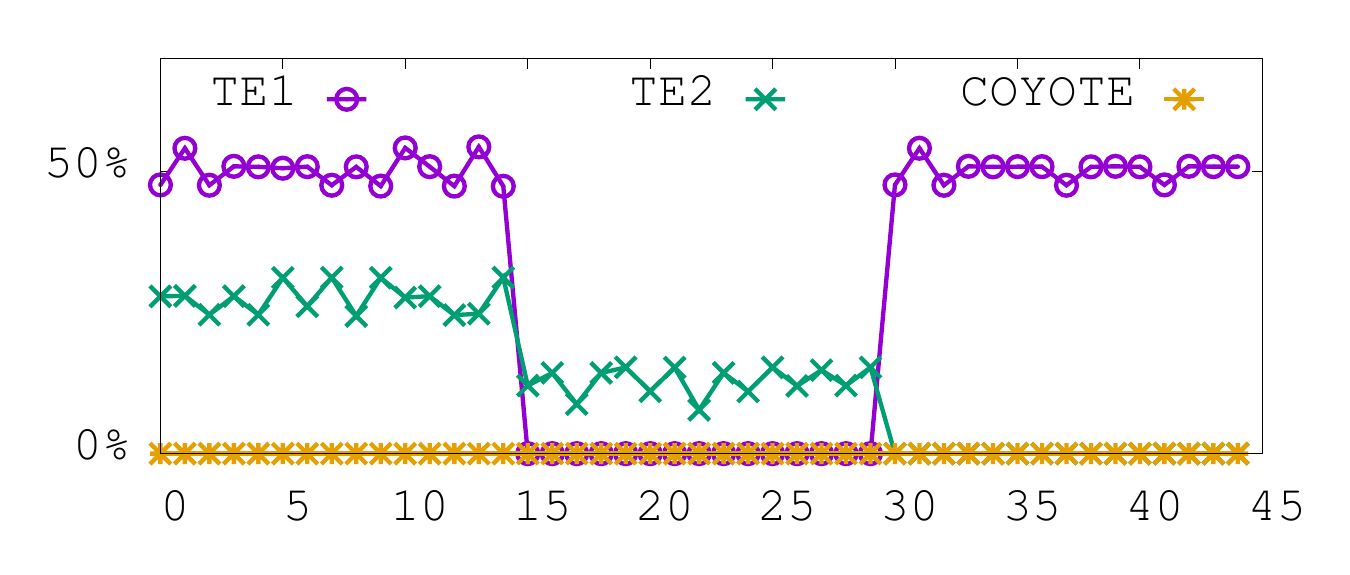}%
		\label{subfig:mininet-example-results}}%
	\caption{Mininet topology (a) and packet drop rate (b).}
	\label{fig:mininet-example-all}
\end{figure}

Consider the example in Fig~\ref{fig:mininet-example}: a target node $t$ advertises two IP prefixes $t_1$ and $t_2$ and two sources, $s_1$ and $s_2$, generate traffic destined for these IP prefixes. As in traditional TE with ECMP, the network operator must use the same forwarding DAG for each destination, this forces either $s_1$ or $s_2$ to route all of its traffic only on the direct path to the destination. Thus, three forwarding DAGs are possible: (1) both $s_1$ and $s_2$ route all traffic on their direct paths to $t$ (TE1), (2) $s_1$ equally splits its traffic between $t$ and $s_2$, and $s_2$ forwards all traffic on its direct link to $t$ (TE2), and (3) same as the previous option, but $s_1$ and $s_2$ swap roles (TE3). 

We evaluate these three TE configurations in mininet with links of bandwidth 1Mbps. We measure the cumulative packet drop rate towards two IP destinations, $t_1$ and $t_2$, for three 15-seconds-long traffic scenarios, where traffic is UDP generated with \texttt{iperf3} and units are in Mbps: $(s_1-t_1,s_2-t_2)=(0,2)$, $(s_1-t_1,s_2-t_2)=(1,1)$, $(s_1-t_1,s_2-t_2)=(2,0)$. 

Fig~\ref{subfig:mininet-example-results} plots the results of this experiment for each of the TE schemes, described above (excluding TE3, which is symmetric to TE2). The x-axis is time (in seconds) and the y-axis is the measured packet loss rate, i.e., the ratio of traffic received to traffic sent (observe that sent traffic is 30 megabits in all scenarios). During the first 15 seconds the experiment emulates the first traffic scenario described above, in the next 15 seconds the second traffic scenario is emulated, and in the last 15 seconds the third scenario is emulated.

Observe that each of the TE schemes (TE1-3) achievable via traditional TE with ECMP leads to a significant packet-drop rate (25\%-50\%) in at least one of traffic scenarios. COYOTE, in contrast, leverages its superior expressiveness to generate different DAGs for each IP prefix destination, as follows: traffic to for destination $t_1$ is evenly split at node $s_1$ and traffic to destination $t_2$ is evenly split at $s_2$. This is accomplished by injecting a lie to $s_2$ so as to attracts half of its traffic to $t_2$ to the $(s_2,s_1)$ link. Consequently, as seen in Fig~\ref{subfig:mininet-example-results}, the rate of dropped packets is significantly reduced.

%% file: 800-tr-related-work.tex
\section{Related Work}\label{sect:related-work}

\noindent{\bf TE with ECMP.} TE with ECMP is today's prevalent approach to TE (see surveys in~\cite{1039866, 4483669}). Consequently, this has been the subject of extensive research and, in particular, selecting good link weights for ECMP TE has received much attention~\cite{DBLP:conf/infocom/FortzT00, 916782, 1424035, fortz-thorup:jsac-2002, fortz-thorup:increasing, Altin:2012:OOR:2345370.2345375, chiesa2014traffic}. To handle uncertainty about traffic demand matrices and variation in traffic, past studies also examined the optimization of ECMP configuration with respect to multiple expected demand matrices~\cite{fortz-thorup:jsac-2002, fortz-thorup:increasing, ericsson:genetic-ospf}, or even with no knowledge of the demand matrices~\cite{4032716}. Unfortunately, while careful optimizations of ECMP configuration can be close-to-optimal in some networks~\cite{fortz-thorup:jsac-2002}, this approach is fundamentally plagued by the intrinsic limitations of ECMP, specifically, routing only on shortest paths and equally splitting traffic at each hop, and can hence easily result in poor network performance. Worse yet, this scheme suffers from inherent computational intractability, as shown in~\cite{DBLP:conf/infocom/FortzT00, chiesa2014traffic}. 

\vspace{0.05in}\noindent{\bf Lying for more expressive OSPF-ECMP routing.} The first technique to approximate unequal splitting through ECMP via the introduction of virtual links was introduced by Nemeth et al. in~\cite{nemeth:networking_2013} (see also \cite{Zhou:2014:WWC:2592798.2592803}). \cite{nemeth:networking_2013}, however, was still limited to shortest-path routing and, consequently, coarse-grained traffic flow manipulation. Recently, Fibbing~\cite{Vissicchio:2014:SLL:2670518.2673868,Vissicchio:2015:CCO:2785956.2787497} showed how any set of destination-based forwarding DAGs can be generated through the injection of fake nodes and links into the underlying link-state protocol (e.g., OSPF).

\vspace{0.05in}\noindent\noindent{\bf Adaptive TE schemes.} One approach to overcoming ECMP's limitations is dynamically adapting the routing of traffic in response to changes in traffic conditions as in, e.g.,~\cite{DBLP:conf/infocom/FortzT00}. Adaptive schemes, however, typically require frequently gathering fairly accurate information about  demand matrices, potentially require new routing or measurement infrastructure, and can be prone to routing instability~\cite{bertsekas:dynamic}, slow convergence, packet reordering, and excess control plane burden \cite{Curtis:2011:DSF:2018436.2018466} (especially in the presence of failures). COYOTE, in contrast, reflects the exact opposite approach: optimizing the \emph{static} configuration of traffic flow so as to simultaneously achieve good network performance with respect to \emph{all}, even adversarially chosen,  demand matrices within specified ``uncertainty bounds''.

\vspace{0.05in}\noindent{\bf Demands-oblivious routing.} A rich body of literature on algorithmic theory investigates so-called ``(demand-)oblivious routing''~\cite{Racke:2008:OHD:1374376.1374415, 4032716, Altin:2012:OOR:2345370.2345375}. Breakthrough algorithmic results by R\"{a}cke established that the static (non-adaptive) routing can be optimized so as to be within an $O(\log n)$ factor from the optimum (demands-aware) routing with respect to \emph{any} combination of  demand matrices \cite{Racke:2008:OHD:1374376.1374415}. Applegate and Cohen~\cite{4032716} showed that when applied to actual (ISP) networks, such demand-oblivious routing algorithms yield remarkably close-to-optimal performance. Kulfi~\cite{kulfi} uses semi-oblivious routing to improve TE in wide-area networks. Unfortunately, all the above demand-oblivious algorithms involve forwarding packets based on both the source and destination, these immediately hit a serious deployability barrier in traditional IP networks (e.g., due to extensive tunneling~\cite{916782}). COYOTE, in contrast, is restricted to OSPF-based destination-based routing, and so tackles inherently different (and new) algorithmic challenges and techniques, as discussed in Sect.~\ref{sec:challenges} and~\ref{sec:design}.

%% file: 900-tr-conclusions.tex
\section{Conclusion}
\label{sec:conc}

We presented COYOTE, a new OSPF-ECMP-based TE scheme that efficiently utilizes the network even with little\slash no knowledge of the traffic demand matrices. We showed that COYOTE significantly outperforms today's prevalent TE schemes while requiring no changes whatsoever to routers. We view COYOTE as an important additional step in the recent exploration~\cite{Vissicchio:2014:SLL:2670518.2673868,Vissicchio:2015:CCO:2785956.2787497} of how SDN functionality can be accomplished without changing today's networking infrastructure. We next discuss important directions for future research.

To efficiently utilize the network in an OSPF-ECMP-compatible manner, COYOTE leveraged new algorithmic insights about destination-based oblivious routing. We believe that further progress on optimizing such routing configurations is key to improving upon COYOTE. Two interesting research questions in this direction: (1) We showed in Section~\ref{sec:challenges} that computing the optimal oblivious IP routing configuration is NP-hard. Can the optimal configuration be provably \emph{well-approximated}? (2) COYOTE first computes a forwarding DAG rooted in each destination node and then computes traffic splitting ratios within these DAGs. The latter computation involves nontrivial optimizations, e.g., via geometric programming, yet, it remains unclear whether traffic splitting within a given set of DAGs is, in fact, efficiently and optimally solvable.


\section*{Acknowledgements}
We thank the anonymous reviewers of the CoNEXT PC and Walter Willinger for their valuable comments.
We thank Francesco Malandrino for useful discussions about the geometric programming approach, and Olivier Tilmans and Stefano Vissicchio for guiding us through the Fibbing code. This research is (in part) supported by European Union's Horizon 2020 research and innovation programme under the ENDEAVOUR project (grant agreement 644960). The 1st and 3rd authors are supported by the Israeli Center for Research Excellence in Algorithms. The 2nd author is with the Department of Telecommunications and Media Informatics, Budapest University of Technology and Economics. 

%% file: 000-abstract.tex
\begin{abstract}
To optimize the flow of traffic in IP networks, operators do traffic engineering (TE), i.e.,  tune routing-protocol parameters in response to traffic demands. TE in IP networks typically involves configuring static link weights and splitting traffic between the resulting shortest-paths via the Equal-Cost-MultiPath (ECMP) mechanism. Unfortunately, ECMP is a notoriously cumbersome and indirect means for optimizing traffic flow, often leading to poor network performance. Also, obtaining accurate knowledge of traffic demands as the input to TE is elusive, and traffic conditions can be highly variable, further complicating TE. We leverage recently proposed schemes for increasing ECMP's expressiveness via carefully disseminated bogus information ("lies") to design COYOTE, a readily deployable TE scheme for robust and efficient network utilization. COYOTE leverages new algorithmic ideas to configure (static) traffic splitting ratios that are optimized with respect to all (even adversarially chosen) traffic scenarios within the operator's "uncertainty bounds". Our experimental analyses show that COYOTE significantly outperforms today's prevalent TE schemes in a manner that is robust to traffic uncertainty and variation. We discuss experiments with a prototype implementation of COYOTE.
\end{abstract}

%% file: 100-intro.tex
\section{Introduction}\label{sect:intro}

To adapt the routing of traffic to the demands network operators do traffic engineering (TE), i.e., tune routing-protocol parameters so as to influence how traffic flows in their networks~\cite{1039866, 4483669, Curtis:2011:DSF:2018436.2018466}. Today's prevalent scheme for TE within an organizational IP network is based on configuring static link-weights into shortest-path protocols such as OSPF~\cite{rfc2328} and splitting traffic between the resulting shortest-paths via ECMP \cite{rfc2992}. Traditional TE with ECMP significantly constrains both route-computation and traffic splitting between multiple paths in two crucial ways: (1) traffic from a source to a destination in the network can only flow along the shortest paths between them (for the given configuration
of link weights), and (2) traffic splitting between multiple paths (if multiple shortest paths exist) can only be done in very specific manners (see Section~\ref{sec:motivating-example} for an illustration). 

ECMP's lack of expressiveness makes traffic engineering with ECMP a notoriously hard task that often results in poor performance. Indeed, not only does ECMP's inflexibility imply that traffic flow might be arbitrarily far from the global optimum~\cite{fortz-thorup:increasing}, but even choosing ``good'' link weights for TE with ECMP is infeasible in general~\cite{chiesa2014traffic}. Beyond ECMP's deficiencies, today's dominant TE schemes also suffer from other predicaments, e.g., obtaining an accurate view of traffic demands so as to optimize TE is elusive, as most networks lack the appropriate measurement infrastructure. Also, traffic can be highly variable and routing configurations that are good with respect to one traffic scenario can be bad with respect to another. We thus seek a TE scheme that is backwards compatible with legacy routing infrastructure (i.e., OSPF and ECMP), yet \emph{robustly} achieves high performance even under uncertain or variable traffic conditions.

\vspace{0.05in}\noindent{\bf Introducing COYOTE: optimized, OSPF/ECMP-compatible TE.} We leverage recently introduced approaches for enriching ECMP's expressiveness without changing router hardware\slash software to design COYOTE (COmpatible Yet Optimized TE). Recent studies show that by injecting ``lies'' into OSPF-ECMP (specifically, information about fake links and nodes), OSPF and ECMP can support much richer traffic flow configurations~\cite{Vissicchio:2014:SLL:2670518.2673868,Vissicchio:2015:CCO:2785956.2787497}. We exploit these developments to explore how OSPF-ECMP routing can be extended to achieve consistently high performance even under great uncertainty about the traffic conditions and high variability of traffic. To accomplish this, COYOTE relies on new algorithmic ideas to configure (static) traffic splitting ratios at routers\slash switches that are optimized with respect to \emph{all} (even \emph{adversarially} chosen) traffic scenarios within operator-specified ``uncertainty bounds''. 

We view COYOTE as an important additional step in the recent exploration of how SDN-like functionality can be accomplished without changing today's networking infrastructure (see~\cite{Vissicchio:2014:SLL:2670518.2673868,Vissicchio:2015:CCO:2785956.2787497}). Our experimentation with COYOTE on real network topologies shows that COYOTE indeed consistently and robustly achieves good performance even with very limited (in fact, sometimes even no) knowledge about the traffic demands and, in particular, exhibits significantly better performance than (optimized) traditional TE with ECMP. Our experiments with a prototype implementation of COYOTE also demonstrate its performance benefits. We briefly discuss below the algorithmic challenges facing the design of COYOTE and how these are tackled.

\vspace{0.05in}\noindent{\bf New algorithmic framework: destination-based oblivious routing.} A rich body of literature in algorithmic theory investigates ``(traffic-demands-)oblivious routing''~\cite{Racke:2008:OHD:1374376.1374415, 4032716, Altin:2012:OOR:2345370.2345375}, i.e., how to compute provably good routing configurations with limited (possibly even no) knowledge of the traffic demands. Past studies~\cite{4032716, kulfi} show that, even though lacking accurate information about the traffic demands,  demands-oblivious routing algorithms yield remarkably close-to-optimal performance on real-world networks. Unfortunately, the above-mentioned algorithms involve forwarding packets based on both source and destination and are so inherently incompatible with des\-ti\-na\-tion-based routing via OSPF-ECMP. In addition, realizing these schemes in practice entails either excessive use of (e.g., MPLS) tunneling\slash tagging in traditional IP networks~\cite{4032716,applegate-thorup-03}, or the ubiquitous deployment of per-flow routing software-defined networking infrastructure~\cite{openflow}.

Our design of COYOTE relies on a novel algorithmic framework for demands-oblivious IP routing. We initiate the study of optimizing oblivious routing under the restriction that forwarding is destination-based. In light of the recent progress on enhancing OSPF-ECMP's expressiveness, we view the algorithmic investigation of destination-based oblivious routing as an important and timely research agenda. We take the first steps in this direction. Our first result establishes that, in contrast to unconstrained oblivious routing, computing the optimal destination-based oblivious routing configuration is computationally intractable. We show how, via the decomposition of this problem into sub-problems that are easier to address with today’s mathematical toolkit, and by leveraging prior research, good routing configurations can be generated. We regard our algorithmic results along these lines as a first, yet promising, step en route to better TE in IP networks, and leave the reader with many interesting open questions in Section~\ref{sec:conc}.


%% file: 300-overview-and-design.tex
\section{COYOTE: Overview and Design}
\label{sec:overview}

 We next motivate COYOTE through a simple example, present the algorithmic challenges facing COYOTE's design, and explain how these are tackled. 


  \input{200-motivating-example}

\if \TR 0
\vfill\eject
\fi
\subsection{Challenges}

Realizing hop-by-hop des\-ti\-na\-tion-based traffic engineering boils down to computing, for each destination, (1) a Directed Acyclic Graph (DAG) rooted in that destination (so as to guarantee loop-free packet forwarding) along which traffic is to be forwarded, and (2) the splitting of traffic within each DAG (c.f., Fig.~\ref{fig:opt-COYOTE}). Thus, the optimization problem that underlies COYOTE is the following: compute per-des\-ti\-na\-tion DAGs and traffic splitting ratios so as to minimize the worst-case link utilization (also referred to as congestion in TE literature~\cite{1039866, 4483669}) across \emph{all} possible traffic scenarios (within the operator-specified uncertainty bounds). Our first (and negative) result shows that this is, in fact, intractable. 
\if \TR 0
    The proof is omitted due to space constraints and is available in~\cite{coyote-tr}.
\fi




\newcommand{\theohardness}{Given a capacitated network graph $G=(V,E)$ and a set ${\cal D}$ of possible traffic demands between nodes, computing the optimal combination of DAGs and traffic splitting ratios with respect to minimizing the worst-case link utilization is \nphard. \if \TR 1 The problem is \nphard\ even if ${\cal D}$ consists of only two possible traffic demands between two source vertices and a single target vertex . \fi}

\begin{theorem}\label{thm:np-complete}
\theohardness
\end{theorem}


Hence, efficiently computing the \emph{optimal} selection of DAGs \emph{and} in-DAG traffic splitting ratios is beyond reach. We next describe how COYOTE's design addresses this challenge. COYOTE's flow-computation decomposes the task of computing destination-based oblivious routing configurations into two sub-problems, and tackles each independently. First, COYOTE applies a simple heuristic to compute destination-oriented DAGs. Then, COYOTE optimizes in-DAG traffic splitting ratios through a combination of optimization techniques, including iterative geometric programming. We show in Section~\ref{sec:eval} that COYOTE's routing algorithm empirically exhibits good network performance.

\subsection{COYOTE Design}\label{sect:COYOTE-design}

Figure~\ref{fig:COYOTE-architecture} presents an overview of the COYOTE architecture. COYOTE gets as input the (capacitated) network topology and the so-called ``uncertainty bounds'', i.e., for every two nodes (routers) in the network, $i$ and $j$, a real-valued interval $[d_{ij}^{min},d_{ij}^{max}]$, capturing the operator's uncertainty about the traffic demand from $i$ to $j$ or, alternatively, the potential variability of traffic. COYOTE then uses this information first to compute a forwarding DAG rooted in each destination node, and then to optimize traffic splitting ratios within each DAG. Lastly, the outcome of this computation is converted into OSPF configuration by injecting ``lies'' into routers. We next elaborate on each of these components.

	\begin{figure}
	  \centering
	\begin{small}
	\tikzset{
	>=stealth',
	  invis/.style={
	    draw=none,
	    text centered, 
	    align=center,
	    on chain},
	  punktchain/.style={
	    rectangle, 
	    rounded corners, 
	    align=center,
	    draw=black, thin,
	    inner sep=3pt,
	    minimum height=3em, 
	    text centered, 
	    on chain},
	  line/.style={draw, thin, <-},
	  every join/.style={->, thin,shorten >=1pt},
	}
	  \begin{tikzpicture}
	    [scale=.2,node distance=1.2em, start chain=going right,] %
	     \scriptsize 
	    \node[invis, join] (input) {demands\\uncertainty\\bounds \&\\topology}; %
	    \node[punktchain, join] (dag) [align=center]{DAG\\construction}; %
	    \node[punktchain, join] (gp) {Traffic\\splitting\\ratio\\calculation};%
	    \node[punktchain, join] (fibbing) {OSPF\\translation}; %
	    \node[invis, join] (output) {OSPF\\messages}; %
	  \end{tikzpicture}
	\end{small}%
	  \caption{COYOTE architecture.}%
	  \label{fig:COYOTE-architecture}
	\end{figure}

\vspace{0.05in}\noindent{\bf Computing DAGs.} Theorem~\ref{thm:np-complete} implies that computing DAGs so as to support optimal routing (through the appropriate in-DAG traffic splitting) is intractable. In COYOTE, DAGs rooted in different destinations are not coupled in any way, allowing network operators to specify any set of DAGs. We show in Section~\ref{sec:eval}, however, that COYOTE significantly outperforms TE with ECMP even when the underlying DAGs are selected with the following simple heuristic: (1) compute, for each destination, the shortest-path DAG rooted in that destination when the link weights are the inverse capacities, and then (2) augment each DAG with additional links by orienting each link that does not appear in the shortest-path DAG towards the incident node that is closer to the destination, breaking ties lexicographically (suppose that the nodes are numbered). Revisiting our running example in Fig.\ref{fig:sample-net}, observe that while the shortest-path DAG rooted at $t$ does \emph{not} contain link $(s_2,v)$ if all links have the same weight, the augmented forwarding DAGs will also utilize this link (in some direction).

\vspace{0.05in}\noindent{\bf Computing traffic splitting ratios for each DAG.} The second fundamental building block of COYOTE is an algorithm that receives as input a set of per-destination DAGs and optimizes traffic splitting within these DAGs, with the objective of minimizing the worst-case congestion (link utilization) over a given set of possible traffic demands. Whether this problem can be solved optimally in a computationally-efficient manner remains an open question (see Section~\ref{sec:conc}. This seems impossible within the familiar mathematical toolset of TE, namely, integer and linear programming. We found that a different approach is, however, feasible: casting the optimization problem as a geometric program (in fact, a mixed linear-geometric program \cite{boyd:geometric_prog}).

Stating COYOTE's traffic splitting optimization as a geometric program is not straightforward and involves careful application of various techniques (convex programming, monomial approximations, LP duality). We provide an intuitive exposition of some of these ideas below using the running example in Fig.~\ref{fig:COYOTE-overview} 
\if \TR 1
  (more details in Appendix~\ref{sect:app-gp}).
\else
  (more details in in~\cite{coyote-tr}).
\fi

Again, $s_1$ and $s_2$ send traffic to $t$, let the DAG for $t$ be as in Fig. \ref{fig:opt-COYOTE}, and suppose that the capacity on links $(s_1, s_2)$, $(s_1, v)$, and $(s_2, v)$ is infinite (that is, arbitrarily large) and on $(s_2, t)$ and $(v, t)$ is $1$.  We are given a set of possible traffic demands $\{d_{s_1}, d_{s_2}\}$ for the two users and our goal is to find the traffic splitting ratios $\phi$ so that the worst-case link utilization across all demands is minimized. A simplified mathematical program for this problem would take the following form (see explanations below): %
\allowdisplaybreaks %
\begin{align}
  &\qquad \qquad \qquad \min \alpha\label{eq:mp-obj}\\
  &\!\!\!\frac{d_{s_1}  \phi(s_1, s_2)  \phi(s_2, t) + d_{s_2}  \phi(s_2, t)}{\text{capacity}(s_2, t)} \le \alpha \quad \  \forall d_{s_1}, d_{s_2}\label{eq:link_b}\\
  &\!\!\!\begin{aligned}
      \frac{d_{s_1} (1 - \phi(s_1, s_2) \phi(s_2, t)) + d_{s_2}\left(1-\phi(s_2, t)\right)}{\text{capacity}(v,t)} \le \alpha\\
            \forall d_{s_1}, d_{s_2}\label{eq:link_c}
    \end{aligned}\ %
\end{align}
\vspace{-1.3em}%

The objective is to minimize $\alpha$, which represents worst-case link utilization, i.e., the load (flow divided by capacity) on the most utilized link across all the admissible traffic demands. Each variable $\phi(x,y)$ denotes the fraction of the incoming flow at vertex $x$ that is routed on link $(x,y)$. Constraints~\eqref{eq:link_b} and~\eqref{eq:link_c} force $\alpha$ to be at least the value of the link utilization of links $(s_2,t)$ and $(s_1,s_2)$, respectively. For the sake of simplicity, we do not show the link utilization constraints for the remaining links.  Now, consider constraint~\eqref{eq:link_b} for link $(s_2,t)$. Observe that from user $s_1$ the fraction of traffic sent through $(s_2,t)$ equals the fraction of $s_1$'s traffic through $(s_1,s_2)$ (i.e., $\phi(s_1,s_2)$) times the fraction sent through $(s_2,t)$ by $s_2$ (i.e., $\phi(s_2,t)$). The fraction of $s_2$'s traffic through $(s_2,t)$ is simply $\phi(s_2,t)$. Accordingly the total flow on $(s_2,t)$ equals $d_{s_1} \cdot \phi(s_1,s_2) \cdot \phi(s_2,t) + d_{s_2} \cdot \phi(s_2,t)$. Hence, the link utilization of $(s_2,t)$ is this expression divided by the capacity of $(s_2,t)$, and the corresponding constraint \eqref{eq:link_b} requires that this utilization be at most $\alpha$ for \emph{all} demands $d_{s_1}, d_{s_2}$. Constraint \eqref{eq:link_c} states the same for link $(v,t)$, where the fraction of traffic sent by $s_1$ ($s_2$) to $t$ through $(v,t)$ is equal to $1$ minus the fraction of flow sent from $s_1$ ($s_2$) to $t$ through $(s_2,t)$.

Two difficulties with these constraints immediately arise: one is that it is \emph{universally quantified} over an entire set of traffic demands, possibly of infinite cardinality, and the other is that it involves a \emph{product of unknowns}, namely, $\phi(s_1,s_2) \cdot \phi(s_2,t)$, and such products do not fit into the framework of standard linear and integer programming. For a discrete demand set we can handle the first problem by stating \eqref{eq:link_b} and \eqref{eq:link_c} for each individual demand. Otherwise (if the set of demands is of infinite size) the elegant dualization technique from~\cite{4032716} can be used.  To handle the second issue, however, we need a small trick from geometric programming~\cite{boyd:geometric_prog}.  Let $d_{s_1}=1$ and $d_{s_2}=1$ and consider constraint \eqref{eq:link_b}:
\begin{displaymath}
  \phi(s_2,t) + \phi(s_1,s_2) \cdot \phi(s_2,t) \le \alpha \enspace .
\end{displaymath}

Now, substitute for new variables, $\widetilde{\phi}(s_1,s_2) = \log \phi(s_1,s_2)$ and $\widetilde{\phi}(s_2,t) = \log \phi(s_2,t)$, and take the logarithm of both sides:
\begin{displaymath}
  \log\left(e^{\widetilde{\phi}(s_2,t)} + e^{\widetilde{\phi}(s_1,s_2) + \widetilde{\phi}(s_2,t)} \right)\le \log \alpha \enspace .
\end{displaymath}

This constraint is now a logarithm of a sum of exponentials of linear functions and so is convex, opening the door to using standard convex programming. 
Our implementation uses a convex program based on the above ideas and others delicate techniques to compute the traffic splitting ratios. 
\if \TR 0
  The reader is referred to our technical report for a detailed explanation~\cite{coyote-tr}.
\else
  See Appendix~\ref{sect:app-gp} for a detailed explanation.
\fi

\vspace{0.05in}\noindent{\bf Translation to OSPF-ECMP configuration.} As explained above, using OSPF and ECMP for TE constrains the flow of traffic in two significant ways: (1) traffic only flows on shortest-paths (induced from operator specified link weights), and (2) traffic is split equally between multiple next-hops on shortest-paths to a destination. Recent studies show how OSPF-ECMP's expressiveness can be significantly enhanced by effectively deceiving routers. Specifically, Fibbing~\cite{Vissicchio:2014:SLL:2670518.2673868,Vissicchio:2015:CCO:2785956.2787497} shows how \emph{any} set of per-destination forwarding DAGs can be realized by introducing fake nodes and virtual links into an underlying link-state routing protocol, thus overcoming the first limitation of ECMP. \cite{nemeth:networking_2013} shows how ECMP's equal load balancing can be extended to much more nuanced traffic splitting by setting up virtual links alongside existing physical ones, thus relaxing the second of these limitations.

We revisit our running example to show how COYOTE exploits these techniques. Consider Fig.~\ref{fig:COYOTE-fibbing}. Inserting a fake advertisement at $s_1$ into the OSPF link-state database can ``deceive'' $s_1$ into believing that, besides its available shortest paths via $s_2$ and $v$ , destination $t$ is also available via a third, ``virtual'' forwarding path. The forwarding adjacency in the fake advertisement is mapped to $s_2$, so that out of $s_1$'s three next-hops to $t$ node $s_2$ will appear \emph{twice} while $v$ only appears once. Consequently, the traffic is \emph{effectively} split between $s_2$ and $v$ in a ratio \nicefrac{2}{3} to \nicefrac{1}{3}. Beyond changing how traffic is split \emph{within} a given shortest-path DAG, as illustrated in Fig.~\ref{fig:COYOTE-fibbing}, fake nodes\slash links can be injected into OSPF so to as change the forwarding DAGs themselves at the per-IP-destination-prefix granularity, as shown in~\cite{Vissicchio:2015:CCO:2785956.2787497}. COYOTE leverages the techniques in~\cite{Vissicchio:2015:CCO:2785956.2787497} and in~\cite{nemeth:networking_2013} to carefully craft ``lies'' so as to generate the desired per-destination forwarding DAGs and approximate the optimal traffic splitting ratios with ECMP.  Section~\ref{sec:eval} that highly optimized TE is achievable even with the introduction of few virtual nodes and links. 

%% file: 200-motivating-example.tex
\if \TR 0 
  \subsection{Motivating Example}
  \label{sec:motivating-example}
\else 
    \section{Motivating Example}
    \label{sec:motivating-example}
\fi

\begin{figure}
  \centering
  \subfloat[][]{%
      \begin{tikzpicture}
        [scale=.18, baseline=(s1.base), minimum size=15,inner sep=2pt, %
        node/.style={anchor=center,circle,draw=black,font=\normalsize}] {
          \node (s1) at (0,0) [node] {$s_1$};%
          \node (s2) at (8,5) [node] {$s_2$};%
          \node (v) at (8,-5) [node] {$v$};%
          \node (t) at (16,0) [node] {$t$};%
        
          \path (s1) edge (s2) edge (v); %
          \path (s2) edge (v) edge (t); %
          \path (v) edge (t); %
        };
      \end{tikzpicture} %
    \label{fig:sample-net}}%
   \hskip1.3em%
    \subfloat[][]{%
      \begin{tikzpicture}
        [scale=.18, baseline=(s1.base), minimum size=15,inner sep=2pt,    %
        node/.style={anchor=center,circle,draw=black,font=\normalsize}] {
          \node (s1) at (0,0) [node] {$s_1$};%
          \node (s2) at (8,5) [node] {$s_2$};%
          \node (v) at (8,-5) [node] {$v$};%
          \node (t) at (16,0) [node] {$t$};%
          
          \path (s1) edge (s2) edge (v); %
          \path (s2) edge (v) edge (t); %
          \path (v) edge (t); %

          \draw[->,>=latex,dashed] ($ (s2) + (1,-1.5) $) -- ($(v)+(1,1.5)$); %
          \node at ($(s2) + (2,-3.8)$) [node,rectangle,draw=none,font=\small] {$\nicefrac{1}{2}$}; %

          \draw[->,>=latex,dashed] ($ (s2) + (2,0) $) -- ($(t) + (-1.25,1.95)$); %
          \node at ($(s2) + (3.25,0.15)$) [node,rectangle,draw=none,font=\small] {$\nicefrac{1}{2}$}; %

          \draw[->,>=latex,dashed] ($ (s1) + (1,1.75) $) -- ($(s2) + (-1.65,.25)$); %
          \node at ($(s1) + (.5,3.15)$) [node,rectangle,draw=none,font=\small] {$\nicefrac{1}{2}$}; %

          \draw[->,>=latex,dashed] ($ (s1) + (1,-1.75) $) -- ($(v) + (-1.65,-0.25)$); %
          \node at ($(s1) + (.5,-3.15)$) [node,rectangle,draw=none,font=\small] {$\nicefrac{1}{2}$}; %

          \draw[->,>=latex,dashed] ($ (v) + (2,0) $) -- ($(t) + (-1.25,-1.95)$); %
          \node at ($(v) + (3.25,-.15)$) [node,rectangle,draw=none,font=\small] {$1$}; %
	};	
      \end{tikzpicture}
    \label{fig:opt-ecmp}}%
  \vspace{-1.5em}%
  \newline%
  \subfloat[][]{%
      \begin{tikzpicture}
        [scale=.18, baseline=(s1.base), minimum size=15,inner sep=2pt,    %
        node/.style={anchor=center,circle,draw=black,font=\normalsize}] {
          \node (s1) at (0,0) [node] {$s_1$};%
          \node (s2) at (8,5) [node] {$s_2$};%
          \node (v) at (8,-5) [node] {$v$};%
          \node (t) at (16,0) [node] {$t$};%
          
          \path (s1) edge (s2) edge (v); %
          \path (s2) edge (v) edge (t); %
          \path (v) edge (t); %

          \draw[->,>=latex,dashed] ($ (s2) + (1,-1.5) $) -- ($(v)+(1,1.5)$); %
          \node at ($(s2) + (2,-3.8)$) [node,rectangle,draw=none,font=\small] {$\nicefrac{1}{2}$}; %

          \draw[->,>=latex,dashed] ($ (s2) + (2,0) $) -- ($(t) + (-1.25,1.95)$); %
          \node at ($(s2) + (3.25,0.15)$) [node,rectangle,draw=none,font=\small] {$\nicefrac{1}{2}$}; %

          \draw[->,>=latex,dashed] ($ (s1) + (1,1.75) $) -- ($(s2) + (-1.65,.25)$); %
          \node at ($(s1) + (.5,3.15)$) [node,rectangle,draw=none,font=\small] {$\nicefrac{2}{3}$}; %

          \draw[->,>=latex,dashed] ($ (s1) + (1,-1.75) $) -- ($(v) + (-1.65,-0.25)$); %
          \node at ($(s1) + (.5,-3.15)$) [node,rectangle,draw=none,font=\small] {$\nicefrac{1}{3}$}; %

          \draw[->,>=latex,dashed] ($ (v) + (2,0) $) -- ($(t) + (-1.25,-1.95)$); %
          \node at ($(v) + (3.25,-.15)$) [node,rectangle,draw=none,font=\small] {$1$}; %
        };
      \end{tikzpicture}
    \label{fig:opt-COYOTE}}%
  \subfloat[][]{%
      \begin{tikzpicture}
        [scale=.18, baseline=(s1.base), minimum size=15,inner sep=2pt,    %
        node/.style={anchor=center,circle,draw=black,font=\normalsize},
        fakenode/.style={anchor=center,draw=red,dashed,very thick,circle,font=\normalsize}, %
        ] {

          \node (fake-t) at (2,6) [fakenode] {\textcolor{red}{$t$}};%

          \node (s1) at (0,0) [node] {$s_1$};%
          \node (s2) at (8,5) [node] {$s_2$};%
          \node (v) at (8,-5) [node] {$v$};%
          \node (t) at (16,0) [node] {$t$};%
          
          \path (s1) edge (s2) edge (v); %
          \path (s2) edge (v) edge (t); %
          \path (v) edge (t); %

          \draw[->,>=latex,dashed] ($ (s2) + (1,-1.5) $) -- ($(v)+(1,1.5)$); %
          \node at ($(s2) + (2,-3.8)$) [node,rectangle,draw=none,font=\small] {$\nicefrac{1}{2}$}; %

          \draw[->,>=latex,dashed] ($ (s2) + (2,0) $) -- ($(t) + (-1.25,1.95)$); %
          \node at ($(s2) + (3.25,0.15)$) [node,rectangle,draw=none,font=\small] {$\nicefrac{1}{2}$}; %

          \draw[->,>=latex,dashed,color=red,thick] ($ (s1) + (0,1.75) $) .. controls ($ (s1) + (-2.5,3.75) $) .. ($(fake-t) + (-1.65,-.25)$); %
          \node at ($(s1) + (-3,3.15)$) [node,rectangle,draw=none,font=\small] {$\nicefrac{1}{3}$}; %

          \draw[->,>=latex,dashed,color=red,thick] ($ (fake-t) + (1.50,-0.25) $) -- ($(s2) + (-1.65,.55)$); %
          \node at ($(s1) + (-3,3.15)$) {}; %

          \draw[->,>=latex,dashed] ($ (s1) + (1,1.75) $) -- ($(s2) + (-1.65,.25)$); %
          \node at ($(s1) + (1.5,3.15)$) [node,rectangle,draw=none,font=\small] {$\nicefrac{1}{3}$}; %

          \draw[->,>=latex,dashed] ($ (s1) + (1,-1.75) $) -- ($(v) + (-1.65,-0.25)$); %
          \node at ($(s1) + (.5,-3.15)$) [node,rectangle,draw=none,font=\small] {$\nicefrac{1}{3}$}; %

          \draw[->,>=latex,dashed] ($ (v) + (2,0) $) -- ($(t) + (-1.25,-1.95)$); %
          \node at ($(v) + (3.25,-.15)$) [node,rectangle,draw=none,font=\small] {$1$}; %
        };
      \end{tikzpicture}
    \label{fig:COYOTE-fibbing}}%
   \caption{A sample network: (a) topology with unit capacity links; (b) 
   per-destination ECMP routing (oblivious performance ratio $\nicefrac{3}{2}$); (c) COYOTE (oblivious performance ratio $\nicefrac{4}{3}$); and (d) COYOTE implementation with a fake node inserted at $s_1$ for realizing the required splitting ratio.}
  \label{fig:COYOTE-overview}
\end{figure}

          





\if \TR 1
We next motivate COYOTE through a simple example. 
\fi
Consider the toy example in Fig.~\ref{fig:sample-net}. Two network users, $s_1$ and $s_2$, wish to send traffic to target $t$. Suppose that each user is expected to send between 0 and 2 units of flow and each link is of capacity 1. Suppose also that the network operator is oblivious to the actual traffic demands or, alternatively, that traffic is variable and user demands might drastically change over time. The operator aims to provide robustly good network performance, and thus has an ambitious goal: configuring OSPF-ECMP routing parameters so as to minimize link over-subscription across \emph{all} possible combinations of traffic demands within the above-specified uncertainty bounds.

Consider first the traditional practice of splitting traffic equally amongst the next-hops on shortest-paths to the destination (i.e., traditional TE with ECMP, see Fig.~\ref{fig:opt-ecmp}),  where the shortest paths towards $t$ are depicted by dashed arrows labelled with the resulting flow splitting ratios. Observe that if the actual traffic demands are $2$ and $0$ for $s_1$ and $s_2$, respectively, routing as in Fig.~\ref{fig:opt-ecmp} would result in link (over-)utilization that is $\nicefrac{3}{2}$ higher than that of the optimal routing of these specific demands (which can send all traffic without exceeding any link capacity). Specifically, routing as in Fig.~\ref{fig:opt-ecmp} would result in $\nicefrac{3}{2}$ units of traffic traversing link $(v,t)$, whereas the total flow could be optimally routed without at all exceeding the link capacities by equally splitting it between paths $(s_1\ s_2\ t)$ and $(s_1\ v\ t)$. One can actually show that this is, in fact, the \emph{best} guarantee achievable for this network via traditional TE with ECMP, i.e., for \emph{any} choice of link weights, equal splitting of traffic between shortest paths would result in link utilization that is $\nicefrac{3}{2}$ higher than optimal for \emph{some} possible traffic scenario. Can we do better?

We show that this is indeed possible if more flexible traffic splitting than that of traditional TE with ECMP is possible. One can prove that for \emph{any} traffic demands of the users, per-destination routing as in Fig.~\ref{fig:opt-COYOTE} results in a maximum link utilization at most $\nicefrac{4}{3}$ times that of the optimal routing
\if \TR 1\footnote{In fact, even the routing configuration in Fig.~\ref{fig:opt-COYOTE} is not optimal in this respect. Indeed, COYOTE's optimization techniques, discussed in Section~\ref{sect:COYOTE-design}, yield configurations with better guarantees (see Appendix~\ref{sect:app-golden-ratio}).}.
 \else
 \footnote{In fact, even the routing configuration in Fig.~\ref{fig:opt-COYOTE} is not optimal in this respect. Indeed, COYOTE's optimization techniques, discussed in Section~\ref{sect:COYOTE-design}, yield configurations with better guarantees (see~\cite{coyote-tr}).}.
\fi 
We explain later how COYOTE realizes such uneven per-destination load balancing without any modification to legacy OSPF-ECMP.

%% file: 400-evaluation.tex
\section{Evaluation}
\label{sec:eval}

\begin{figure*}[!ht]
	\centering
	\begin{minipage}{\linewidth}
		\centering
		\frame{\includegraphics[width=.9\columnwidth]{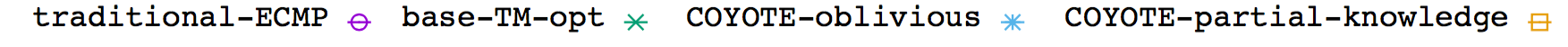}}
		\caption*{}
		\label{fig:leged}
	\end{minipage}
	
	\vspace{-.2in}
	
	\begin{minipage}{.32\linewidth}
		\centering
		\includegraphics[width=\columnwidth]{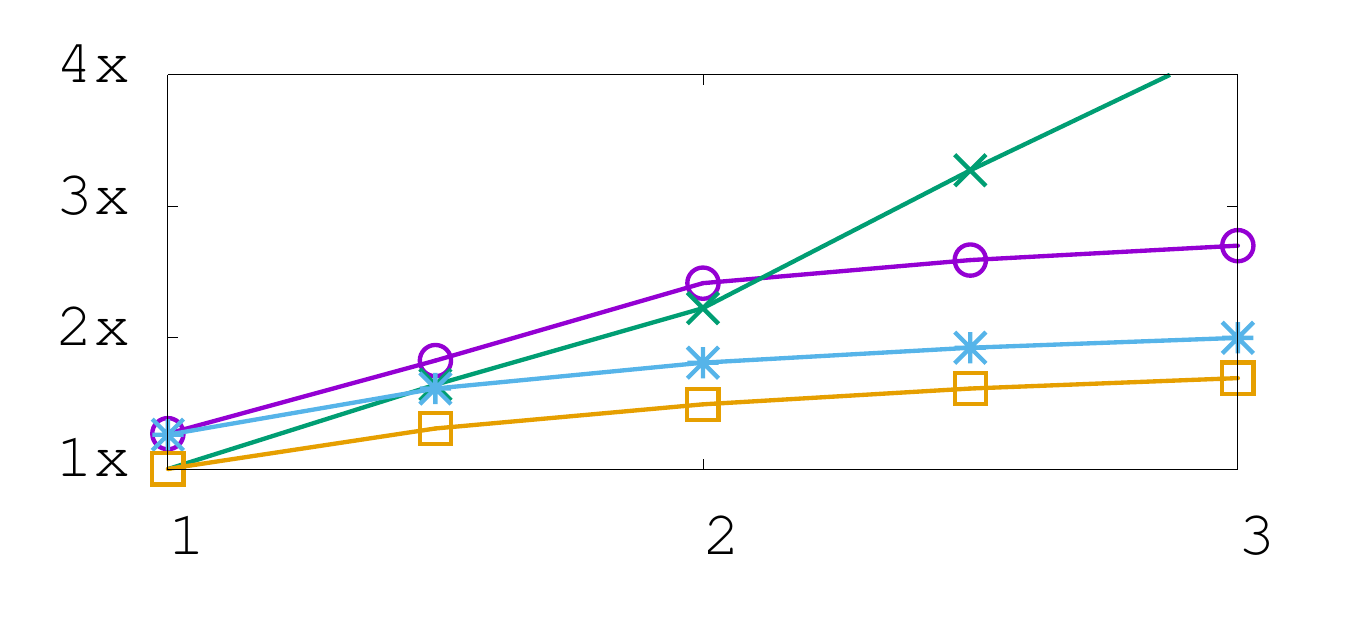}
		\caption{Geant, gravity model.}
		\label{fig:gravity-augmented-shortest-path-dag-geant}
	\end{minipage}
	\begin{minipage}{.32\linewidth}
		\centering
		\includegraphics[width=\columnwidth]{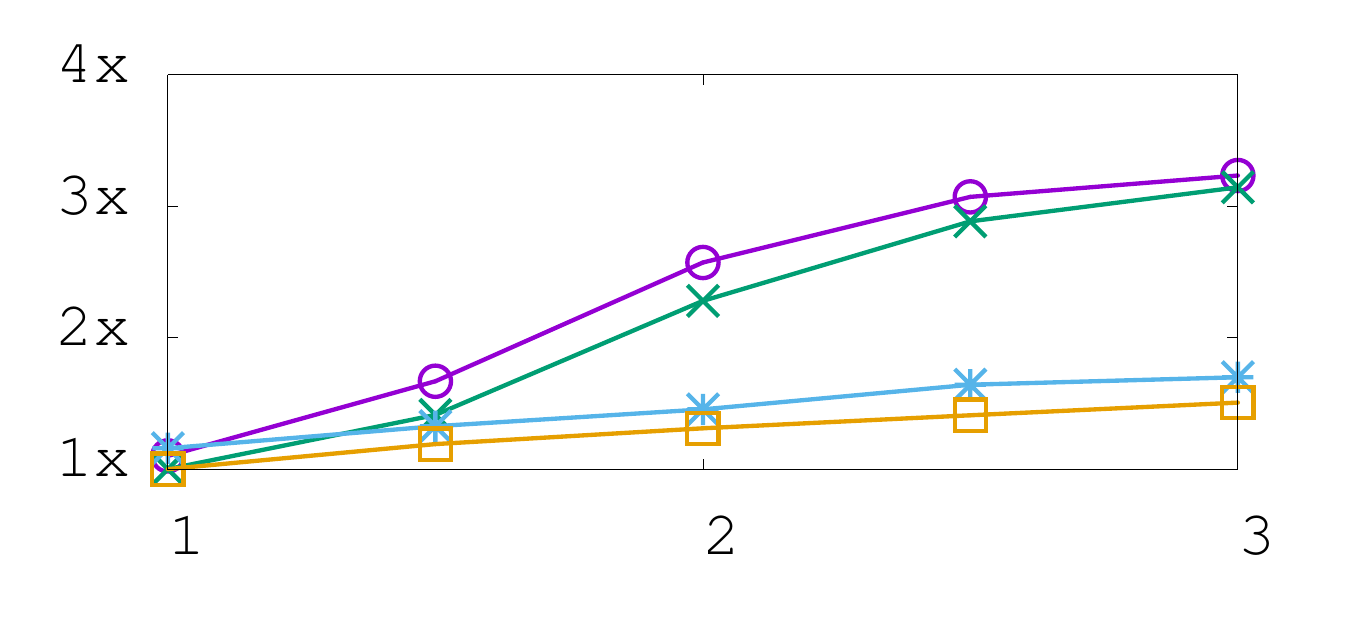}
		\caption{Digex, gravity model.}
		\label{fig:gravity-augmented-shortest-path-dag-digex}
	\end{minipage}
	\begin{minipage}{.32\linewidth}
		\centering
		\includegraphics[width=\columnwidth]{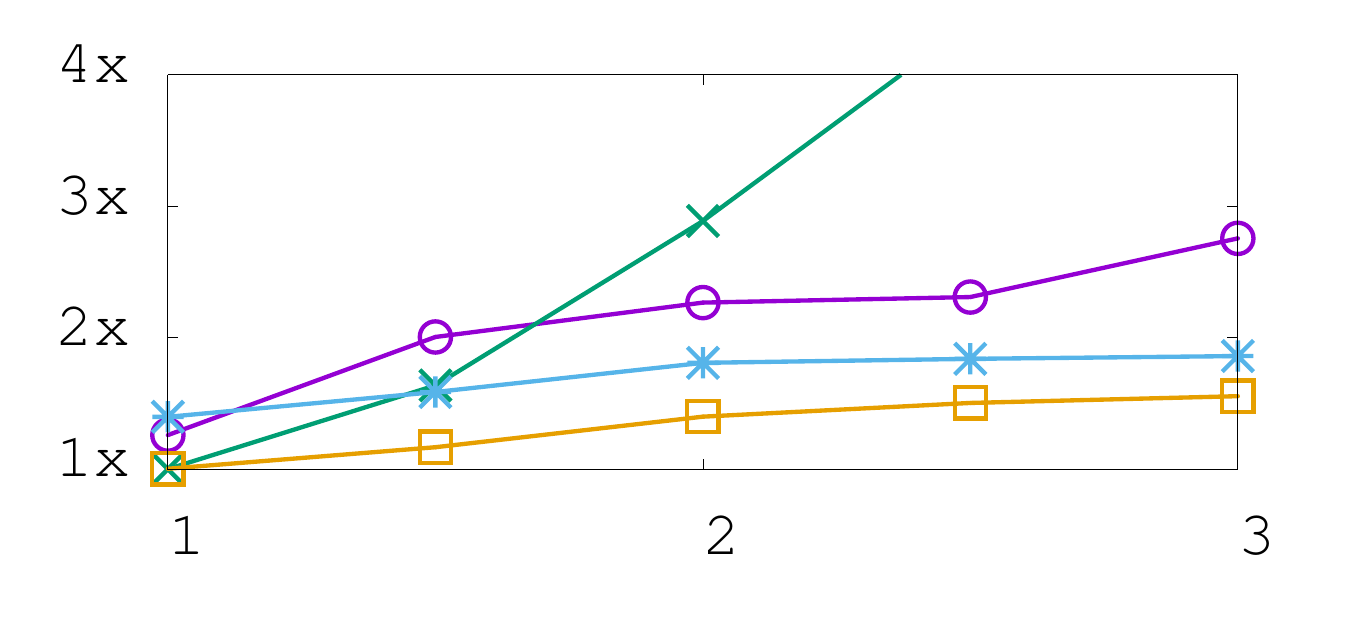}
		\caption{AS 1755, bimodal model}
		\label{fig:bimodal-augmented-shortest-path-dag}
	\end{minipage}
	
	\centering
	
	\begin{minipage}{.32\linewidth}
		\centering
		\includegraphics[width=\columnwidth]{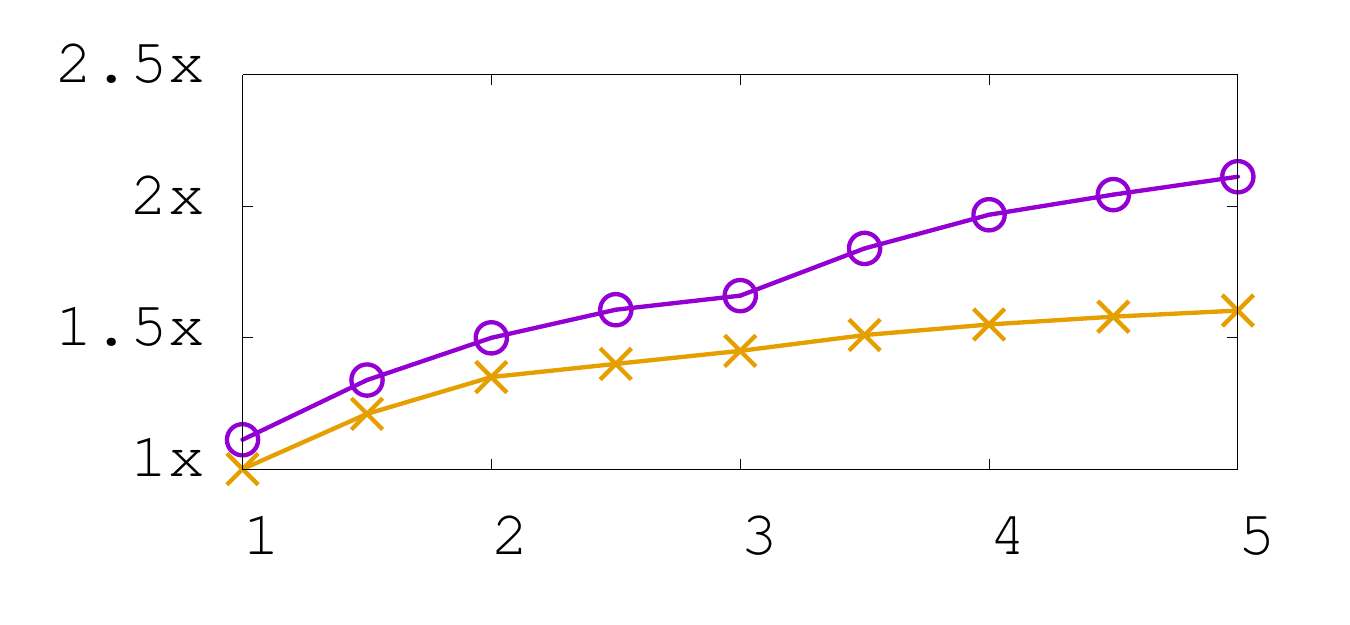}
		\caption{Abilene, optimized-ECMP.}   
		\label{fig:obliviuos-ecmp}
	\end{minipage}
	\begin{minipage}{.32\linewidth}
		\centering
		\includegraphics[width=\columnwidth]{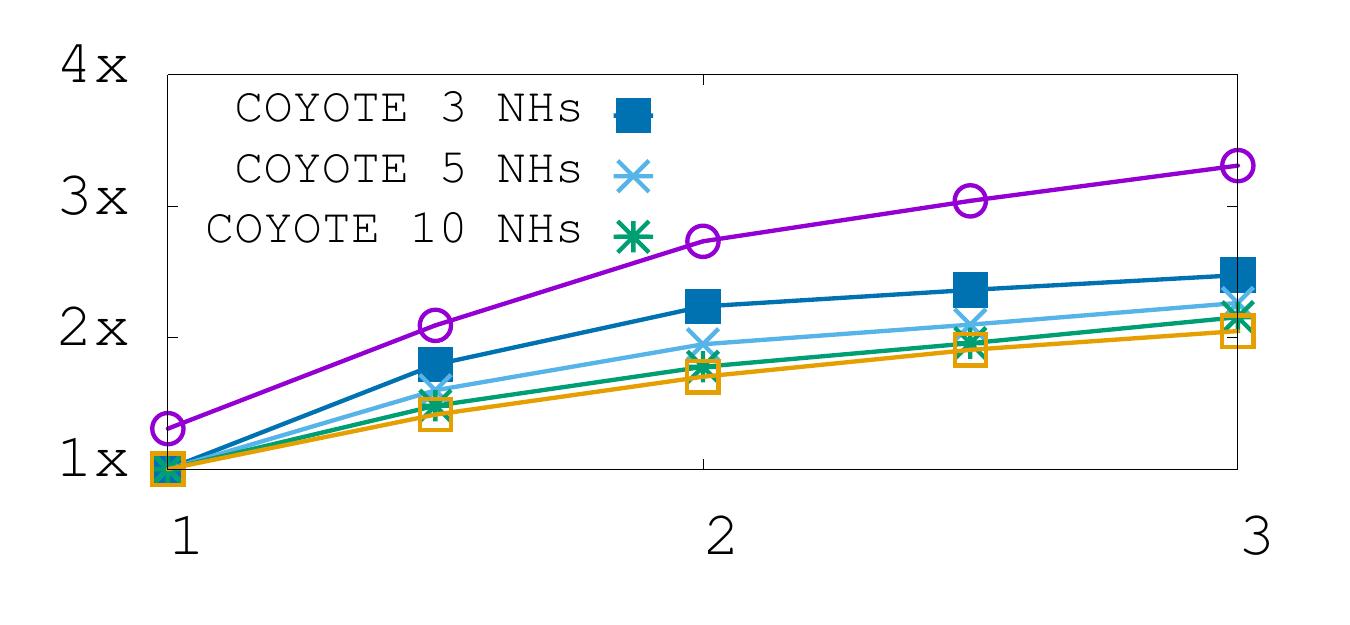}
		\caption{Approximation.}
		\label{fig:approximation}
	\end{minipage}
	\begin{minipage}{.32\linewidth}
		\centering
		\includegraphics[width=\columnwidth]{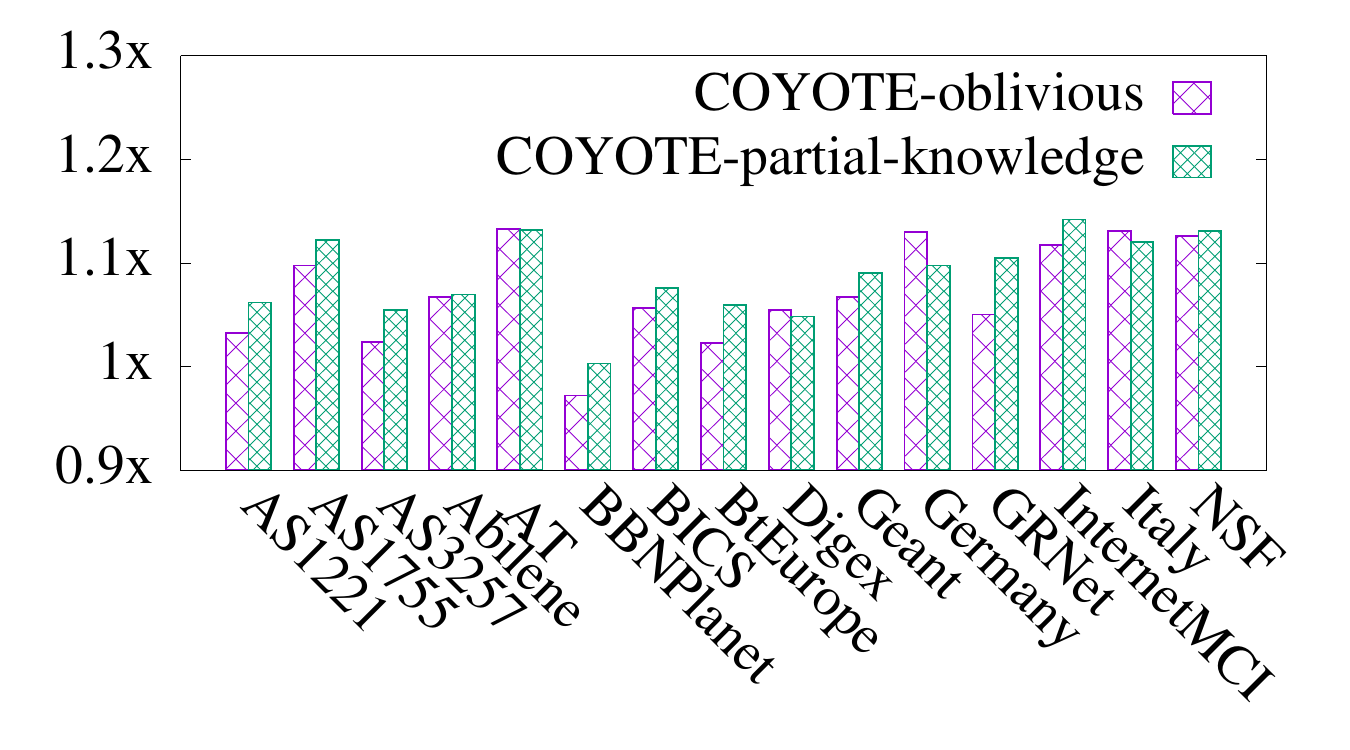}
		\caption{Average stretch.}
		\label{fig:stretch}
	\end{minipage}

\end{figure*}

We experimentally evaluate COYOTE in order to quantify its performance benefits and its robustness to traffic uncertainty and variation. Importantly, our focus is solely on destination-based TE schemes (i.e., TE schemes that can be realized via today's IP routing). We show below that COYOTE provides significantly better performance than ECMP even when \emph{completely} oblivious to the traffic demands. Also, COYOTE's increased path diversity does not come at the cost of long paths: the paths computed by COYOTE are on average only a factor of 1.1 longer than \ecmp's. We also discuss experiments with a prototype implementation of COYOTE.

While the reader might think that COYOTE's performance benefits over traditional TE with ECMP are merely a byproduct of its greater flexibility in selecting DAGs and in traffic splitting, our results show that this intuition is, in fact, false. Specifically, we show that, similarly to unconstrained (i.e., source and destination based) oblivious routing~\cite{4032716}, even the \emph{optimal} routing with respect to \emph{estimated} traffic demands, which can unevenly split traffic, fares much worse than COYOTE if the \emph{actual} traffic demands are not very ``close'' to the estimated demands. Hence, COYOTE's good performance should be attributed not only to its expressiveness but also, in large part, to its built-in algorithms for optimizing performance in the presence of uncertainty, as discussed in Section~\ref{sect:COYOTE-design}.



\subsection{Simulation Framework} 

We use the set of $16$ backbone Internet topologies from the Internet Topology Zoo (ITZ) archive~\cite{topology-zoo} to assess the performance of COYOTE and \ecmp. When available, we use the link capacities provided by ITZ. Otherwise, we set the link capacities to be inversely-proportional to the ITZ-provided \ecmp weights (in accordance with the Cisco-recom\-mend\-ed default OSPF link configuration~\cite{cisco-link-weights}). When neither \ecmp link weights nor capacities are available we use unit capacities and link weights.
%
We evaluate COYOTE against ECMP using two simple DAG-construction heuristics: (1) constructing augmented shortest-path DAGs as explained in Section~\ref{sec:overview} with the link weights provided in the ITZ archive, and (2) optimizing the link weights using the local greedy algorithm of Altin et al~\cite{Altin:2012:OOR:2345370.2345375}. In both cases, we augment the shortest path DAGs as explained in Section~\ref{sec:overview}.

To compute COYOTE's in-DAG traffic splitting ratios (see Section~\ref{sec:overview}), we use AMPL~\cite{ampl} as the problem formulation language and MOSEK~\cite{mosek}, a non-linear convex optimization solver. The running time with our current single-threaded proof-of-concept implementation ranges from few minutes (for small networks) to few days (for large networks). 

While we are currently working on improving the running times, we would like to point out that the computation of the in-DAG traffic splitting ratios needs only be performed
once or on a daily/weekly-base, as routing in COYOTE is not dynamically adjusted, and that routing configurations for failure scenarios (e.g., every single link/node failure) can be precomputed.

We measure performance in terms of the worst-case link utilization (referred to as ``congestion'' in TE literature~\cite{1039866, 4483669}), i.e., the performance of (multicommodity) flow of traffic $f$ is $\max_{l}\frac{f(l)}{c_l}$, where $f(l)$ is the flow traversing link $l$ and $c_l$ is link $l$'s capacity.

\subsection{Network Performance}

We compare COYOTE to ECMP for both DAG-construction heuristics described above and for two types of base traffic demands: (1) \emph{gravity}~\cite{Roughan:2002:EMB:637201.637213}, where the amount of flow sent from router $i$ to router $j$ is proportional to the product of $i$'s total outgoing capacity and $j$'s total outgoing capacity, and (2) \emph{bimodal}~\cite{Medina:2002:TME:964725.633041}, where a small fraction of all pairs of routers exchange large quantities of traffic, and the other pairs send small flows.

We first present our results with respect to the ITZ~\cite{topology-zoo} link weights and an ideal version of COYOTE capable of arbitrarily fine-grained traffic splitting. We then show that a close approximation of the optimal splitting ratios can be obtained with the introduction of a limited number of additional virtual links.
Fig.~\ref{fig:gravity-augmented-shortest-path-dag-geant} and Fig.~\ref{fig:gravity-augmented-shortest-path-dag-digex} describe our results for two networks (Geant and Digex, respectively), the gravity model, and augmented shortest path DAGs based on the ITZ link weights. The x-axis represents the ``uncertainty margin'': let $d_{i,j}$ be the amount of flow from router $i$ to router $j$ in the base traffic demands (namely, gravity), a margin of uncertainty of $x$ means that the actual flow from $i$ to $j$ can be any value between $\frac{d_{i,j}}{x}$ and $x\cdot d_{i,j}$. We increase the uncertainty margin in increments of $0.5$ from $1$ (no uncertainty whatsoever) to $3$ (fairly high uncertainty). The y-axis specifies how far the computed solution is from the \emph{demands-aware} optimum within the same DAGs. 

We plot four lines, corresponding to the performance of four different protocols: (1) traditional TE with ECMP, (2) the optimal demands-aware routing for the base gravity model (with no uncertainty), which can be obtained with linear programming techniques~\cite{multicommodityflow-lp}, (3) COYOTE (oblivious) with traffic splitting optimized with respect to \emph{all} possible traffic demands (i.e., assuming \emph{nothing} about the demands), (4) COYOTE (partial-knowledge) optimized with respect to the traffic demands within the uncertainty margin. Observe that both variants of COYOTE provide significantly better performance than TE with ECMP and, more surprisingly, both COYOTE and (sometimes) ECMP outperform the optimal base routing, whose performance quickly degrades even with little demands uncertainty. Our results thus show that COYOTE's built-in robustness to traffic uncertainty, in the form of optimization under specified uncertainty margins, indeed leads to superior performance in the face of inaccurate knowledge about the traffic demands or, alternatively, variable traffic conditions.
\if \TR 1
Table~\ref{tab:evaluation} shows  the extensive results of COYOTE for all the analyzed topologies, except BBNPlanet and Gambia, which are almost a tree topology.
\input{950-appendix-evaluation}
\fi

We observe the same trends when the base traffic demands are sampled from the bimodal model, as shown in Fig.~\ref{fig:bimodal-augmented-shortest-path-dag}.



We now discuss our results with respect to our second DAG-construction heuristic, which is based on the heuristic of Altin et al.~\cite{Altin:2012:OOR:2345370.2345375} for oblivious \ecmp routing configuration. Specifically, \cite{Altin:2012:OOR:2345370.2345375} presents a heuristic that starts from the link weights provided by the ITZ dataset, and applies a local-search heuristic that greedily changes one link weight if this change improves the worst-case \ecmp link utilization across all the admissible traffic demands. We use the output of this procedure as the ECMP configuration and augment it with additional links to obtain COYOTE's per destination DAGs, as explained in Section~\ref{sec:overview}. Fig.~\ref{fig:obliviuos-ecmp} presents a comparison of COYOTE and \ecmp using the bimodal model as the base traffic demands. We use the above heuristic to compute, for each uncertainty margin in the range $1-5$, increasing in $0.5$ increments, the (traditional) ECMP configuration and COYOTE DAGs with respect to the bimodal-based traffic demands. We then compare the worst-case link utilization of the two, again, normalized by the \emph{demands-aware} optimum within the same (augmented) DAGs. We note that \ecmp is, on average, almost 80\% times further away from the optimum than COYOTE.

\vspace{0.05in}\noindent\textbf{Approximating the optimal traffic splitting.} We evaluated above COYOTE under the assumption that arbitrarily fine-grained traffic splitting is achievable, yet in practice, the resolution of traffic splitting is derived from the number of virtual links introduced. Clearly, an excessive number of virtual links should be avoided for at least two reasons: (a) each virtual next-hop is installed into the finite-sized Forwarding Information Base (FIB), and (b) injecting additional information into OSPF comes at the cost of additional computational overhead. Our results, illustrated in Fig.~\ref{fig:approximation} for AS 1755 network's topology (all other topologies exhibit the same trend), show that even with just $3$ additional virtual links per router interface, COYOTE achieves a 50\% improvement over traditional TE with ECMP. We observe that with $10$ virtual links the computed routing configuration closely approximates the ideal solution.

\vspace{0.05in}\noindent\textbf{Average path lengths.}
COYOTE augments the shortest path DAG with additional links so as to better utilize the network. Consequently, traffic can potentially traverse longer paths. We show, however, that COYOTE's increased path redundancy does not come at the expense of long paths. Specifically, the average stretch (increase in length) of the paths in COYOTE is typically bounded within a 10\% factor with respect to the OSPF/ECMP paths. Fig.~\ref{fig:stretch} plots the average stretch across all pairs for a margin of $2.5$. Similar results are obtained for all different margins between $1$ to $5$. Observe that the DAGs computed by COYOTE rely on shortest-path computation with respect to the link weights, whereas the stretch is measured in terms of the number of hops. Thus, it is possible for the stretch to be less than 1, as is the case, e.g., for BBNPlanet. 

\section{Prototype Implementation}

We implemented and experimented with a prototype of the COYOTE architecture, as described in Section~\ref{sec:overview}. Our prototype extends the Fibbing controller code, written in Python and provided by Vissicchio et al.~\cite{Vissicchio:2015:CCO:2785956.2787497}, and uses the code of Nemeth et al. from~\cite{nemeth:networking_2013} for approximating the splitting ratios. We plan to make our code public in the near future. We next illustrate the benefits of COYOTE over traditional TE, as reflected by an evaluation of our prototype via the mininet~\cite{mininet-hifi} network emulator. 

\begin{figure}
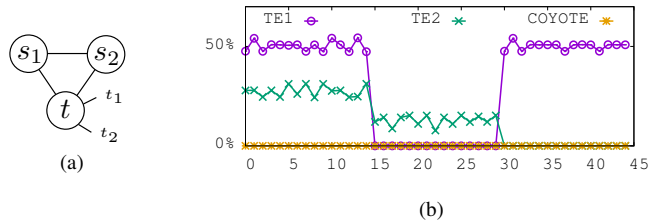

	\centering
	\subfloat[][]{%
		\includegraphics[width=45pt,valign=c]{fig/mininet-instance.pdf}%
		\label{fig:mininet-example}}%
	\hfill%
	\subfloat[][]{%
		\includegraphics[width=180pt,valign=c]{fig/mininet-experiment-2.pdf}%
		\label{subfig:mininet-example-results}}%
	\caption{Mininet topology (a) and packet drop rate (b).}
	\label{fig:mininet-example-all}
\end{figure}

Consider the example in Fig~\ref{fig:mininet-example}: a target node $t$ advertises two IP prefixes $t_1$ and $t_2$ and two sources, $s_1$ and $s_2$, generate traffic destined for these IP prefixes. As in traditional TE with ECMP, the network operator must use the same forwarding DAG for each destination, this forces either $s_1$ or $s_2$ to route all of its traffic only on the direct path to the destination. Thus, three forwarding DAGs are possible: (1) both $s_1$ and $s_2$ route all traffic on their direct paths to $t$ (TE1), (2) $s_1$ equally splits its traffic between $t$ and $s_2$, and $s_2$ forwards all traffic on its direct link to $t$ (TE2), and (3) same as the previous option, but $s_1$ and $s_2$ swap roles (TE3). 

We evaluate these three TE configurations in mininet with links of bandwidth 1Mbps. We measure the cumulative packet drop rate towards two IP destinations, $t_1$ and $t_2$, for three 15-seconds-long traffic scenarios, where traffic is UDP generated with \texttt{iperf3} and units are in Mbps: $(s_1-t_1,s_2-t_2)=(0,2)$, $(s_1-t_1,s_2-t_2)=(1,1)$, $(s_1-t_1,s_2-t_2)=(2,0)$. 

Fig~\ref{subfig:mininet-example-results} plots the results of this experiment for each of the TE schemes, described above (excluding TE3, which is symmetric to TE2). The x-axis is time (in seconds) and the y-axis is the measured packet loss rate, i.e., the ratio of traffic received to traffic sent (observe that sent traffic is 30 megabits in all scenarios). During the first 15 seconds the experiment emulates the first traffic scenario described above, in the next 15 seconds the second traffic scenario is emulated, and in the last 15 seconds the third scenario is emulated.

Observe that each of the TE schemes (TE1-3) achievable via traditional TE with ECMP leads to a significant packet-drop rate (25\%-50\%) in at least one of traffic scenarios. COYOTE, in contrast, leverages its superior expressiveness to generate different DAGs for each IP prefix destination, as follows: traffic to for destination $t_1$ is evenly split at node $s_1$ and traffic to destination $t_2$ is evenly split at $s_2$. This is accomplished by injecting a lie to $s_2$ so as to attracts half of its traffic to $t_2$ to the $(s_2,s_1)$ link. Consequently, as seen in Fig~\ref{subfig:mininet-example-results}, the rate of dropped packets is significantly reduced.

%% file: 500-related-work.tex
\section{Related Work}\label{sect:related-work}

\noindent{\bf TE with ECMP.} TE with ECMP is today's prevalent approach to TE (see surveys in~\cite{1039866, 4483669}). Consequently, this has been the subject of extensive research and, in particular, selecting good link weights for ECMP TE has received much attention~\cite{DBLP:conf/infocom/FortzT00, 916782, 1424035, fortz-thorup:jsac-2002, fortz-thorup:increasing, Altin:2012:OOR:2345370.2345375, chiesa2014traffic}. To handle uncertainty about traffic demands and variation in traffic, past studies also examined the optimization of ECMP configuration with respect to multiple expected traffic demands~\cite{fortz-thorup:jsac-2002, fortz-thorup:increasing, ericsson:genetic-ospf}, or even with no knowledge of the traffic demands~\cite{4032716}. Unfortunately, while careful and delicate optimizations of ECMP configuration can be close-to-optimal in some networks~\cite{fortz-thorup:jsac-2002}, this approach is fundamentally plagued by the intrinsic limitations of ECMP, specifically, routing only on shortest paths and equally splitting traffic at each hop, and can hence easily result in poor network performance. Worse yet, this scheme suffers from inherent computational intractability, as shown in~\cite{DBLP:conf/infocom/FortzT00, chiesa2014traffic}. 

\vspace{0.05in}\noindent{\bf Lying for more expressive OSPF-ECMP routing.} The first technique to approximate unequal splitting through ECMP via the introduction of virtual links was introduced by Nemeth et al. in~\cite{nemeth:networking_2013} (see also \cite{Zhou:2014:WWC:2592798.2592803}). \cite{nemeth:networking_2013}, however, was still limited to shortest-path routing and, consequently, coarse-grained traffic flow manipulation. Recently, Fibbing~\cite{Vissicchio:2014:SLL:2670518.2673868,Vissicchio:2015:CCO:2785956.2787497} showed how any set of destination-based forwarding DAGs can be generated through the injection of fake nodes and links into the underlying link-state protocol (e.g., OSPF).

\vspace{0.05in}\noindent\noindent{\bf Adaptive TE schemes.} One approach to overcoming ECMP's limitations is dynamically adapting the routing of traffic in response to changes in traffic conditions as in, e.g.,~\cite{DBLP:conf/infocom/FortzT00}. Adaptive schemes, however, typically require frequently gathering fairly accurate information about traffic demands, potentially require new routing or measurement infrastructure, and can be prone to routing instability~\cite{bertsekas:dynamic}, slow convergence, packet reordering, and excess control plane burden \cite{Curtis:2011:DSF:2018436.2018466} (especially in the presence of failures). COYOTE, in contrast, reflects the exact opposite approach: optimizing the \emph{static} configuration of traffic flow so as to simultaneously achieve good network performance with respect to \emph{all}, even adversarially chosen, traffic demands within specified ``uncertainty bounds''.

\vspace{0.05in}\noindent{\bf Demands-oblivious routing.} A rich body of literature on algorithmic theory investigates so-called ``(demand-)oblivious routing''~\cite{Racke:2008:OHD:1374376.1374415, 4032716, Altin:2012:OOR:2345370.2345375}. Breakthrough algorithmic results by R\"{a}cke established that the static (non-adaptive) routing can be optimized so as to be within an $O(\log n)$ factor from the optimum (demands-aware) routing with respect to \emph{any} combination of traffic demands \cite{Racke:2008:OHD:1374376.1374415}. Applegate and Cohen~\cite{4032716} showed that when applied to actual (ISP) networks, such demand-oblivious routing algorithms yield remarkably close-to-optimal performance. Kulfi~\cite{kulfi} uses semi-oblivious routing to improve TE in wide-area networks. Unfortunately, all the above demand-oblivious algorithms involve forwarding packets based on both the source and destination, these immediately hit a serious deployability barrier in traditional IP networks (e.g., due to extensive tunneling~\cite{916782}). COYOTE, in contrast, is restricted to OSPF-based destination-based routing, and so tackles inherently different (and new) algorithmic challenges and techniques, as discussed in Sect.~\ref{sec:overview}.

%% file: 600-conclusions.tex
\section{Conclusion}
\label{sec:conc}

We presented COYOTE, a new OSPF-ECMP-based TE scheme that efficiently utilizes the network even with little\slash no knowledge of the traffic demands. We showed that COYOTE significantly outperforms today's prevalent TE schemes while requiring no changes whatsoever to routers. We view COYOTE as an important additional step in the recent exploration~\cite{Vissicchio:2014:SLL:2670518.2673868,Vissicchio:2015:CCO:2785956.2787497} of how SDN functionality can be accomplished without changing today's networking infrastructure. We next discuss two important directions for future research.

\vspace{0.05in}\noindent{\bf Further exploring destination-based oblivious routing.} To efficiently utilize the network in an OSPF-ECMP-compatible manner, COYOTE leveraged new algorithmic insights about destination-based oblivious routing. We believe that further progress on optimizing such routing configurations is key to improving upon COYOTE. We next mention two interesting research questions in this direction: (1) We showed in Section~\ref{sec:overview} that computing the optimal oblivious IP routing configuration is NP-hard. Can the optimal configuration be provably \emph{well-approximated}? (2) COYOTE first computes a forwarding DAG rooted in each destination node, and then computes traffic splitting ratios within each DAG. The latter computation involves nontrivial optimizations, e.g., via iterative geometric programming, yet, it remains unclear whether traffic splitting within a given set of DAGs is, in fact, efficiently and optimally solvable.

\vspace{0.05in}\noindent{\bf Evaluating COYOTE under real-world network conditions.} Our experimental evaluation of COYOTE combined simulations on $16$ backbone Internet topologies from the Internet Topology Zoo (ITZ) archive~\cite{topology-zoo} with small-scale experiments on the mininet network emulator~\cite{mininet-hifi}. An important direction for future research is experimenting with COYOTE on empirically-derived traffic traces, e.g., data collected from the Internet2 Network~\cite{internet2}, and running larger-scale experiments with our prototype implementation of COYOTE on mininet and on SDN-capable network testbeds such as~\cite{ocean}.


\section*{Acknowledgements}
We thank the anonymous reviewers of the CoNEXT PC and Walter Willinger for their valuable comments.
We thank Francesco Malandrino for useful discussions about the geometric programming approach, and Olivier Tilmans and Stefano Vissicchio for guiding us through the Fibbing code. This research is (in part) supported by European Union's Horizon 2020 research and innovation programme under the ENDEAVOUR project (grant agreement 644960). The 1st and 3rd authors are supported by the Israeli Center for Research Excellence in Algorithms. The 2nd author is with the Department of Telecommunications and Media Informatics, Budapest University of Technology and Economics. 

%% file: 950-tr-app-closer-look.tex
\appendix

We present below a more detailed exposition of COYOTE's algorithmic machinery. Recall that COYOTE's traffic flow optimization consists of two steps: (1) computing per-destination DAGs (Section~\ref{sec:computing-dags}), and (2) optimizing traffic splitting \emph{within} these DAGs (Section~\ref{sec:splitting-ratio-overview}). We next dive into the technical details involved in overcoming these challenges. In Appendix~\ref{sec:local-search-dag}, we describe the local search heuristic for computing ``good'' DAGs. We then discuss in-DAG traffic splitting. Specifically, in Appendix~\ref{sec:revising-example} we revisit our running example and show how the optimal traffic splitting ratios can be computed for this specific instance of \obliviousperdestination. Then, in Appendix~\ref{sec:dual-gp} we explain how COYOTE leverages dualization theory and Geometric Programming (GP) to compute in-DAG traffic splitting in general.

\appendices

\section{The Local Search DAG-Generation Algorithm}\label{sec:local-search-dag}

COYOTE utilizes an adaptation of the local search DAG-generation heuristic from~\cite{Altin:2012:OOR:2345370.2345375}. The pseudocode is given in Algorithm~\ref{alg:local-search}.

\begin{algorithm}
  \caption{The Local search DAG generation algorithm.}
  \label{alg:local-search}
  \begin{small}
    \begin{algorithmic}[1]
      \State{\textsc{Input:} graph $G(V, E)$ with link capacities $c: E \mapsto \mathbb{R}$}
      \State{\textsc{Output:} a link cost function $w: E \mapsto \mathbb{N}$}
      \State{$\mathcal{D} \gets \emptyset$}
      \State{$w \gets$ \textsc{InverseCapacity}$(c)$}
      \While {(true)}
      \State{\textbf{for each} $t \in V: DAG_t \gets$ \text{Shortest-Path-First}$(G, w, t)$}\label{alg-line:spf-dag}
      \State{$DM \gets$ \textsc{WorstCaseDm}$(G, \{DAG_t\})$}\label{alg-line:worst-case-tm}
      \State{$\mathcal{D} \gets \mathcal{D} \cup \{DM\}$}\label{alg-line:add-tm}
      \State{\textbf{if} \textsc{MaxLinkUtil}$(G, \{DAG_t\}, \mathcal{D})$ $\le B$ \textbf{then break}}\label{alg-line:exit condition}
      \State{$w \gets$ \textsc{FortzThorup}$(G, \mathcal{D}, c)$}\label{alg-line:fortz-thorup}
      \EndWhile
    \end{algorithmic}
  \end{small}
\end{algorithm}

The algorithm maintains a set $\mathcal{D}$ of ``critical'' demand matrices (DMs) and iteratively tries to find DAGs that, when distributing traffic using ECMP, yield low link utilization across these DMs. COYOTE's non-equal splitting ratios can allow even lower utilization of links. In each iteration, the following steps are executed: (1) Compute shortest-path DAG $DAG_t$ to each destination $t \in V$ for the current link weights $w$ (line~\ref{alg-line:spf-dag}), (2) Compute the DM that produces the highest link utilization over the resulting routing (see~\cite{Altin:2012:OOR:2345370.2345375} for a mathematical program that captures this task), (3) Add this DM to $\mathcal{D}$ (lines~\ref{alg-line:worst-case-tm} and~\ref{alg-line:add-tm}), and (4) Recompute weights $w$ inducing DAGs that are \emph{simultaneously} good with respect to each DM in $\mathcal{D}$ (line~\ref{alg-line:fortz-thorup}). Specifically, use the tabu search technique due to Fortz and Thorup~\cite{fortz-thorup:increasing}, which iteratively tries to reduce utilization at the most congested node by increasing the path diversity locally. The heuristic terminates when maximum link utilization reduces below some pre-configured bound $B$.  

The above heuristic modifies~\cite{fortz-thorup:increasing} and~\cite{Altin:2012:OOR:2345370.2345375} to better fit COYOTE's design, namely, \emph{(i)} the OSPF-TE cost-optimization heuristics of Fortz and Thorup scale link utilizations through a non-linear function $\Phi$ to penalize heavily loaded links whereas our heuristic optimizes for maximum link-utilization (as oblivious routing optimizes for max link-utilization); \emph{(ii)} when optimizing DAGs with respect to multiple DMs, Fortz and Thorup use the average of the scaled link utilizations over the DMs while we do simple maximum; and \emph{(iii)} our local search algorithm is fine-tuned to COYOTE by carefully selecting the parameters governing the heuristic search process.

\section{Revisiting the Running Example}\label{sec:revising-example}

Recall the simple example in Fig.~\ref{fig:COYOTE-overview}. We now show how to to compute its optimal in-DAG traffic splitting ratios $\phi$. The input DAG is depicted by dashed arrows in the figure. 

As discussed in Section~\ref{sec:np-hard}, we can focus, without loss of generality, only on those demand matrices that are non-dominated vertices of the polyhedron representing the set of DMs that can be routed without exceeding the edge capacities. In our example, the set of demand matrices that can be routed without exceeding the edge capacities is $\{(d_{s_1t},d_{s_2t}) | d_{s_1t} + d_{s_2t} = 2\}$, and the only two non-dominated vertices are 
$D_1=\{(d_{s_1t},d_{s_2t})=(2,0)\}$ and $D_2=\{(d_{s_1t},d_{s_2t})=(0,2)\}$.  
Let 
$PERF(e,\phi,\{D_1,D_2\})$ denote the worse-case link utilization of an edge $e$ when using a PD routing $\phi$ to route DMs $D_1$ or $D_2$. Henceforth, for brevity, $\{D_1,D_2\}$ shall remain fixed and is thus omitted from the arguments involving $PERF$.

Given any routing $\phi$, observe that $PERF((s_1,v),\phi)\le PERF((v,t),\phi)$ and $PERF((s_2,v),\phi)\le PERF((v,t),\phi)$ since link $(v,t)$ carries the incoming flows from $(s_1,v)$ and $(s_2,v)$.
We hence restrict our focus to $PERF((s_1,s_2),\phi)$, $PERF((v,t),\phi)$, and $PERF((s_2,t),\phi)$. As for $D_1$, the most utilized edge must be either $(s_1,s_2)$ or $(v,t)$ since $(s_2,t)$ carries no more traffic than $(s_1,s_2)$.
This implies that
\begin{align}
PERF((s_1,s_2),&\phi) \ge 2 \phi(s_1,s_2) \label{eq:s1-s2} \text{ and} \\
PERF((v,t),\phi) & \ge 2(1 - \phi(s_1,s_2)) + 2\phi(s_1,s_2) (1-\phi(s_2,t)) \nonumber \\
& \ge 2(1 - \phi(s_1,s_2) \phi(s_2,t)) \label{eq:v-t}
\end{align}

Regarding DM $D_2$, observe that the most congested edge is either $(s_2,t)$ or $(v,t)$, such that:
\begin{align}
PERF((s_2,t),\phi) & \ge 2 \phi(s_2,t)  \label{eq:s2-t} \\
PERF((v,t),\phi) & \ge 2(1 - \phi(s_2,t)) \label{eq:v-t-2}
\end{align}

As for $PERF((v,t),\phi)$, observe that $ 2(1 - \phi(s_1,s_2) \phi(s_2,t)) \ge  2(1 - \phi(s_2,t))$, for any $0\le \phi(s_2,t) \le 1$, which
means that inequality~(\ref{eq:v-t-2}) is redundant.

Observe that (\ref{eq:s1-s2}) increases w.r.t. $\phi(s_1,s_2)$,~(\ref{eq:s2-t}) increases w.r.t. $\phi(s_2,t)$, while (\ref{eq:v-t}) decreases w.r.t. both $\phi(s_1,s_2)$ and $\phi(s_2,t)$. So, to minimize the worse-case link utilization, inequalities (\ref{eq:s1-s2}), (\ref{eq:s2-t}), and (\ref{eq:v-t}) must be tight in the optimal scenario, i.e., $PERF((s_1,s_2),\phi)=PERF((s_2,t),\phi)=PERF((v,t),\phi)$.
Moreover, inequalities (\ref{eq:s1-s2}) and (\ref{eq:s2-t}) imply that $\phi(s_1,s_2)= \phi(s_2,t)$, which allows us to rewrite (\ref{eq:v-t}) as $PERF((v,t),\phi)=1 - \phi(s_1,s_2)^2$. From (\ref{eq:s1-s2}) and~(\ref{eq:v-t}), we have that $  2\phi(s_1,s_2) = 2(1 - \phi(s_1,s_2)^2) \rightarrow
1 - \phi(s_1,s_2) -\phi(s_1,s_2)^2 =0 $, which is an equation of the second order with solutions $\frac{\sqrt{5}-1}{2}$ and $\frac{\sqrt{5}+1}{2}$.
Only the first of these two solutions (i.e., the inverse of the golden ratio) is feasible in our formulation. The optimal splitting ratios are therefore $ \phi(s_1,s_2)= \phi(s_2,t)= \frac{\sqrt{5}-1}{2}$. Traffic splitting accordingly guarantees that the worse-case link utilization on ${\cal D}$ is never greater than $\sqrt{5}-1\sim 1,23$.

\section{Dualization and Geometric Programming}\label{sec:dual-gp}

We explained in Section~\ref{sec:splitting-ratio-overview} how the optimal traffic splitting ratios \emph{within} a DAG (given as \emph{input}) can be computed is the specific scenario considered. Can the optimal traffic splitting ratios \emph{always} be computed in a computationally-efficient manner? While this remains an important open question (see Section~\ref{sec:conc}), it seems impossible to accomplish within the familiar mathematical toolset of TE, namely, integer and linear programming. 
We found that a different approach for generating good in-DAG traffic splitting is, however, feasible:  casting the optimization problem described in Section~\ref{sec:notation-problem} as a mixed-linear geometric program~\cite{boyd:geometric_prog}). 

{We observed in Section~\ref{sec:splitting-ratio-overview} that two difficulties arise when computing in-DAG traffic splitting over a set of possible DMs: (i) the cardinality of the DMs set can possibly be infinite and (ii) modeling per-destination routing involves a product of unknowns. To address (i), we build upon the standard dualization techniques for efficiently optimizing over infinite sets of DMs. We adapt these techniques to the restriction that routing be destination-based. We refer the reader to~\cite{4032716} for additional details about this approach. To address (ii), we leverage Geometric Programming (GP) for approximating non-convex constraints involving products of unknowns with convex constraints. We refer the reader to~\cite{boyd:geometric_prog} for a detailed exposition of GP. We next dive into the details.

We define each DAG rooted at a vertex $t \in V$ as a set of directed edges $E_t$ and let ${\cal E}=\{E_{t_1},\dots,E_{t_n}\}$ for each $t_i \in V$, with $i=1,\dots,n$.
As observed in Section~\ref{sec:splitting-ratio-overview}, since the performance ratio is invariant to any proportional rescaling of the DMs or link capacities, we can reformulate our optimization problem as follows:  
\begin{align}
& \min  \  \alpha \nonumber\\
&  (\phi,f) \text{ is a PD routing in } {\cal E}  \nonumber\\
& \forall \text{ edges } e=(u,v): \nonumber \\
& ~~\forall \text{ DMs }D \in {\cal D} \text{ with } \lambda>0 \text{ such that:}\nonumber \\
& ~~~~OPTU(D)=1 \text{ and } \forall i,j\ \lambda d_{st}^{min}\le d_{st}\le \lambda d_{st}^{max} : \label{eq:demand-bounds}  \\
& ~~~~~~  \nicefrac{\sum_{(s,t)} d_{st}f_{st}(u)\phi_t(e)}{c(e)} \le \alpha  \label{eq:constraints}
\end{align}

Variable $\lambda$ is used to scale each DM  $D$ in ${\cal D}$ so that $OPTU(D)=1$. In addition, we force each $\phi_t$ to be routed within the given DAGs that are defined in ${\cal E}$.

Recall that when ${\cal D}$ is the set of all possible DMs, the performance ratio is referred to as the {\em oblivious} performance ratio. In this case, for each demand $d_{st}$, we simply replace in~(\ref{eq:demand-bounds}) the $\lambda d_{st}^{min}\le d_{st}\le \lambda d_{st}^{max}$ inequality with $0 \le d_{st}$.

\vspace{.1in}
\noindent\textbf{Reducing the number of constraints via duality transformations. }
To simplify exposition, we first explain in detail how to apply the dualization technique only for the scenario that the demand matrix set is unbounded, i.e., \emph{any} traffic matrix is possible. We will later discuss the scenario that the demand matrix set is constrained. 
Recall the formulation of \obliviousperdestination, as described in Section~\ref{sec:notation-problem}. Observe that the constraints at equation~\eqref{eq:constraints} can be tested by solving, for each edge $e=(u,v)$, the following ``slave Linear Problem (LP)'' and checking if the objective is $\le \alpha$. 
\begin{align}
& \max \nicefrac{\sum_{(st)} d_{st}f_{st}(u)\phi_t(e)}{c_e} \nonumber \\
&  \forall t\in V, \forall s\neq t \in V: \nonumber \\
&\sum_{a\in (OUT(s) \wedge E_t)}g_t(a) -\sum_{a\in (IN(s) \wedge E_t)}g_t(a) - d_{st} \le 0 \label{eq:absolute-flow}\\
& \forall a\in E: \textstyle{\sum_{t\in V}}g_t(a) \le c_a \label{eq:link-constraint-g} \\
& \forall s,t \in V: d_{st}\ge 0, \forall t\in V, \forall a\in E: g_t(a)  \ge 0 \nonumber 
\end{align}

Given a fixed routing $(\phi,f)$, the objective function is maximizing the link utilization of $e$ by exploring the set of demand matrices that can be routed within the link capacities of the given DAGs. Variable $g_t(a)$ is a PD routing that represents the amount of absolute flow to $t$ that is routed through edge $a$. Eq.~\eqref{eq:absolute-flow} captures the standard flow conservation constraints for each vertex of the network, where $d_{st}$ is a variable represeting the $s\rightarrow t$ demand. Eq.~\eqref{eq:link-constraint-g} guarantees that $g$ can be routed within the link capacities.

By applying duality theory to the slave LP, we can describe a set of requirements that must be satisfied by a routing $\phi$ in order to guarantee an oblivious performance ratio $\le \alpha$.\vspace{0.1in}

\begin{theorem}\label{theo:duality}
	A routing $\phi$ has oblivious ratio $r$ if there exist positive weights $\pi_e(h)$ for every pair of edges $e,h$, such that:
	\begin{enumerate}[label={R\arabic*}]
		\item $\sum_{h \in E} \pi_e(h)c_h \le r$, for every edge $e \in E$. 
		\item For every edge $(u,v) \in E$, for every demand $s \rightarrow t$, and for every path $(a_1,a_2,\dots,a_l)$ from $s$ to $t$, where $a_1,\dots,a_l \in E_t$, it holds $f_{st}(u)\phi_t(u,v) \le c_e \sum_{k=1}^{l}\pi_e(a_k) $.
	\end{enumerate}
\end{theorem}
\begin{IEEEproof}
	Our proof is based on applying simple duality theory to the slave LP problem. The two requirements are equivalent to stating that the slave LP has objective $\le r$.
	
	Let $\phi$ be a PD routing and $\pi_e(h)$ be weights satisfying requirements $1$-$3$. Let $D$ be any DM that can be routed within the edge capacities and let $q_{st}(a)$ and $q_{st}(p)$ be the amount of the $s\rightarrow t$ demand that is routed through an edge $a$ and  a path $p=(a_1,\dots,a_l)$ according to any routing that does not exceed the edge capacities. Let $e$ be an edge in $E$ and denote $f_{st}(u)\phi_t(u,v)$ by $l_{s,t}(u,v)$. By first multiplying both sides of R2 by $\bar q_{st}(p)$, we obtain
	%
	$$ l_{s,t}(u,v)\bar q_{st}(p) \le c_e \textstyle{\sum_{k=1}^{l}}\pi_e(a_k)\bar q_{st}(p) $$
	
	\noindent and by then summing over this inequality for all the $s\rightarrow t$ paths $p_i=(a_1^i,\dots,a_{l_i}^i)$ such that each edge on the path is in $E_t$, we obtain
	\begin{align}
	\sum_{p_i} l_{s,t}(u,v)\bar q_{st}(p_i) & \le \sum_{p_i} c_e \sum_{k=1}^{l_i}\pi_e(a_k^i)\bar q_{st}(p_i) \nonumber \\
	l_{s,t}(u,v) \sum_{p_i}\bar q_{st}(p_i) & \le   c_e \sum_{h\in E}\pi_e(h)\sum_{p | p \text{ traverses } h}\bar q_{st}(p) \nonumber \\
	l_{s,t}(u,v)d_{st} & \le  c_e \sum_{h\in E}\pi_e(h)q_{st}(h) \text{.} \nonumber 
	\end{align}

	Now, by summing over all the $(s,t)$ pairs, we get	
	$$\sum_{s,t}l_{s,t}(u,v)d_{st}  \le c_e\sum_{h\in E}\pi_e(h)\sum_{s,t}q_{st}(h)   \le  c_e \sum_{h\in E}\pi_e(h)c_h $$
	\noindent where the last inequality holds since $q$ can be routed without exceeding the edge capacities. Combining the above inequality with R1, we finally obtain
	$$\textstyle{\sum_{s,t}} l_{s,t}(u,v)d_{st} \le  c_e \textstyle{\sum_{h\in E}}\pi_e(h)c_h \le c_e r \text{.}$$
	
	This concludes the statement of the theorem as it shows that using $\phi$ to route any DM that can be routed without exceeding the edge capacities would not cause any edge to be over-utilized by a factor higher than $r$.
\end{IEEEproof}

Based on the requirements of Theorem~\ref{theo:duality}, the \obliviousperdestinationdag formulation can be rewritten as the following Non Linear Problem (NLP). Let, for each edge $e$ and pair of vertices $i,j \in V$, the variable $p_e(i,j)$ be the length of the shortest path from $i$ to $j$ according to the edge weights $\pi_e(h)$ (for all $h \in E$). The introduction of these variables allows us to replace the exponential number of constraints (for all possible paths) in Requirement (2) of Theorem~\ref{theo:duality} with a polynomial number of constraints. The final formulation consists of $O(|V|^2|E|)$ variables and $O(|V||E|^2)$ constraints. 
\begin{align}
& \min  \  \alpha \nonumber \\
&  (\phi,f) \text{ is a PD routing} \nonumber \\
& \forall \text{ edges } e\in E \text{:} \nonumber \\
& ~~\textstyle{\sum_{h \in E}} \pi_e(h)c_h \le r  \label{eq:linear-1} \\
& ~~\forall \text{ pairs } (s,t): \nicefrac{f_{,t}(u)\phi_t(u,v)}{c_e}\le p_e(s,t)  \label{eq:gp-constraint} \\
& ~~\forall i\in V, \forall a=(j,k)\in E_t: \nonumber \\
&~~~~~~\pi_e(a) + p_e(k,i) - p_e(j,i) \ge 0 \label{eq:linear-2} \\
& ~~\forall \ h \in E: \pi_e(h) \ge 0; \forall i,j \in V: p_e(i,i)=0, p_e(i,j)\ge 0 \nonumber
\end{align}

\vspace{.1in}
\noindent\textbf{Geometric Programming transformation. }NLP problems are, in general, hard to optimize. We leverage techniques from Geometric Programming (GP) to tackle this challenge~\cite{boyd:geometric_prog}. A {\em Mixed Linear Geometric Programming} (MLGP) is an optimization problem of the form
\begin{align*}
\text{min }   & f_0(x) + a^T_0 y \\
& f_i(x)+ a^T_i y + d_i \le 1 ,i =1 ,\dots,m\\
& h_j(x)=1 ,j =1 ,\dots,M 
\end{align*}

\noindent where $x$ and $y$ are variables, $f_i(x)$ is a sum of {\em posynomials}, i.e., monomials with positive coefficients, $h_j(x)$ is a monomial, and both $a$ and $d$ are vectors of real numbers. Such problems can be transformed with a simple variable substitution $z=log x$ into convex optimization problems~\cite{boyd:geometric_prog}, thus opening the doors to the usage of efficient solvers such as the Interior Point Method. Since our problem contains some constraints that are not posynomials but rather a sum of monomials with positive and negative coefficients, i.e., a signomial, the Complementary GP technique~\cite{boyd:geometric_prog} is used. This involves iteratively approximating the non-GP constraints around a solution point so that the problem becomes MLGP, solving it efficiently, and repeating this procedure with the new solution point. 
We now show how to transform our original NLP dualized formulation into an iterative MLGP formulation.

Routing variables $f$ and $\phi$ are GP variables as they are multiplied with each other in the definition of PD routing. The remaining variables are linear variables. Constraints~\eqref{eq:linear-1} and~\eqref{eq:linear-2} are linear constraints, while Constraint~\eqref{eq:gp-constraint} is an MLGP constraint. As for the flow conservation constraints defined in Section~\ref{sec:notation-problem} of a PD routing, we note that $f_{st}(v)\ge \sum_{e=(u,v) \in E_t}  f_{st}(u) \phi_t(e)$ is a GP constraint but the splitting ratio constraint $\sum_{(v,u)\in E_t} \phi_t(v,u) \ge 1$ is not and we approximate it using monomial approaximation as follows.

Let ${\cal S}_{vt}(\phi)= \sum_{e=(v,u)\in E_t} \phi_t(e)$, where $\phi$ is an array of all the $\phi_t$ variables in the sum. Let $\phi(i)$ the $i$'th variable in $\phi$. Given a point $\phi_0$, we want to approximate ${\cal S}_{vt}$ with a monomial $
k\prod^n_{i =1}(\phi(i))^{a(i)}$. 
From~\cite{boyd:geometric_prog}, we have that $a(i) = \nicefrac{\phi_0(i)}{\sum_{i} \phi_0(i)}$ and $k= \nicefrac{\sum_{i} \phi_0(i)}{\prod_{i=1}^{n}{(\phi_0(i))^{a(i)}}}$.
Hence, each splitting ratio constraint can be rewritten by GP monomial constraint of the form $ 1\le k\prod^n_{i =1}(\phi(i))^{a(i)}$.

To summarize the iterative phase: given a feasible routing solution $\phi_0$, for each destination $t$, we can compute $a_t(u,v)$ and $k_t$ using the above monomial approximation, which leads to the following formulation:
\begin{align}
 & \min  \  \alpha \nonumber \\
 & \forall s,t,v \in V: \tilde f_{st}(v) \ge \log \textstyle{\sum_{(u,v)\in E_t}} e^{\tilde f_{st}(u) +\tilde \phi_t(u,v) } \nonumber \\
& \forall s,t \in V: \tilde f_{st}(s) \ge 0 \nonumber \\
& \forall v,t \in V:\log k_t +  \sum_{h=(v,u) \in E_t} a_t(h)\tilde\phi_t(h)\ge 0  \nonumber \\
& \forall  e\in E : \nonumber \\
& ~~\textstyle{\sum_{h \in E}} \pi_e(h)c_h \le r \nonumber\\
& ~~\forall (s,t)\in V: e^{\tilde f_{st}(u)+\tilde\phi_t(u,v)}\le c_ep_e(s,t)  \label{eq:replaced-in-margin} \\
& ~~\forall t\in V, \forall a=(j,k)\in E_t:  \pi_e(a) + p_e(k,t) - p_e(j,t) \ge 0 \nonumber \\
& ~~\forall \ h \in E: \pi_e(h) \ge 0; \forall i,j \in V: p_e(i,i)=0 \text { and } p_e(i,j)\ge 0 \nonumber
\end{align}

When the set of admissible DMs is bounded, as in the general problem formulation, a similar dualization technique and MLGP transformation can be applied to the problem. One has to carefully consider the uncertainty bounds constraints during the dualization phase, which will be treated as (mixed) linear constraints during the MLGP transformation. Given a feasible routing solution $\phi_0$, by following the same dualization technique presented in~\cite{4032716} we add constraint $\sum_{s,t} (d_{st}^{max}s_e^+(s,t) - d_{st}^{min}s_e^+(s,t)) \le 0$ to the above formulation and we replace Constraint~\eqref{eq:replaced-in-margin} with $e^{\tilde f_{st}(u)+\tilde\phi_t(u,v)}\le c_ep_e(s,t) + s_e^+(s,t) - s_e^-(s,t)$, where $s_e^-(s,t)\ge 0, s_e^+(s,t)\ge 0$. The resulting formulation can still be solved with any solver implementing the Interior Point Method.
%
